\documentclass[onecolumn,3p]{elsarticle}
\usepackage{amsmath}
\usepackage{amssymb}
\usepackage{booktabs} 
\usepackage{lipsum}
\usepackage{amsfonts}
\usepackage{graphicx}
\usepackage{epstopdf}
\usepackage{algorithmic}
\usepackage[justification=centering]{subcaption}
\usepackage{graphicx}
\usepackage{multicol}
\usepackage{cellspace}
\usepackage{cancel}
\usepackage{multirow}
\usepackage{xcolor}
\usepackage{framed}
\usepackage{varwidth}
\usepackage{algorithm2e}
\allowdisplaybreaks
\renewcommand{\vec}[1]{\boldsymbol{#1}}

\newcommand{\Gray}{\mbox{Gy}}
\newcommand{\PARA}[1]{\mathcal{#1}}
\newcommand{\myfrac}[3][0pt]{\genfrac{}{}{}{}{\raisebox{#1}{$#2$}}{\raisebox{-#1}{$#3$}}}
\newcommand{\sn}[1]{\ensuremath{\mbox{SN}_{#1}}}
\usepackage{amsopn}

\makeatletter
\newcommand*{\addFileDependency}[1]{% argument=file name and extension
  \typeout{(#1)}% latexmk will find this if $recorder=0 (however, in that case, it will ignore #1 if it is a .aux or .pdf file etc and it exists! if it doesn't exist, it will appear in the list of dependents regardless)
  \@addtofilelist{#1}% if you want it to appear in \listfiles, not really necessary and latexmk doesn't use this
  \IfFileExists{#1}{}{\typeout{No file #1.}}% latexmk will find this message if #1 doesn't exist (yet)
}
\makeatother

\usepackage{amsmath}
\usepackage{amssymb}
\usepackage{graphicx}
\usepackage{color}
\usepackage{todonotes}

\newcommand{\M}{\ensuremath{\mathcal{M}}}

\newcommand{\parD}[1]{\ensuremath{\ensuremath{\mathcal{D}}_{#1}}}

\usepackage{cancel}

\usepackage{amssymb}
\journal{ }

\begin{document}

\begin{frontmatter}

\title{Spatio-temporal modelling of phenotypic heterogeneity in tumour tissues and its impact on radiotherapy treatment.}

\author[label4,label1]{Giulia L. Celora}

\author[label1]{Helen M. Byrne}
\author[label2]{P.G. Kevrekidis}
 \fntext[label4]{celora@maths.ox.ac.uk}
%% \cortext[cor1]{}
\address[label1]{Mathematical Institute, University of Oxford, Oxford, UK}
\address[label2]{Department of Mathematics \& Statistics,
University of Massachusetts, Amherst 01003 USA \fnref{label3}}

\begin{abstract}

	We present a mathematical model that describes how tumour heterogeneity evolves in a tissue slice that is oxygenated by a single blood vessel.
	Phenotype is identified with the stemness level of a cell, $s$, that determines its proliferative capacity, 
	apoptosis propensity and response to treatment. Our study is based on numerical bifurcation analysis and dynamical simulations of a system of coupled non-local (in phenotypic ``space'') partial differential equations that links the phenotypic evolution of the tumour cells to local oxygen levels in the tissue. In our formulation, we consider a 1D geometry where oxygen is supplied by a blood vessel located on the domain boundary and consumed by the tumour cells as it diffuses through the tissue. For biologically relevant parameter values, the system exhibits multiple steady states; in particular, depending on the initial conditions, the tumour is either eliminated (``tumour-extinction") or it persists (``tumour-invasion''). We conclude by using the model to investigate tumour responses to radiotherapy (RT), and focus on establishing which RT strategies can eliminate the tumour. Numerical simulations reveal how phenotypic heterogeneity evolves during treatment and highlight the critical role of tissue oxygen levels on the efficacy of radiation protocols that are commonly used clinically. 
\end{abstract}


\end{frontmatter}

\section{Introduction}

In recent years, there has been a paradigm shift in how we understand cancer. New technologies, such as single-cell sequencing, have showcased the complex composition of tumours where cells with different genotypes or phenotypes coexist in the same tissue region~\cite{Black2021,Marusyk2012,Patel2014}. While cancer was initially defined as a genetic disease, recent evidence suggests that phenotypic heterogeneity arises from both genetic and non-genetic sources of variability, such as epigenetic and environmental factors \cite{Black2021,Cajal2020,Marusyk2020}. Viewed through this lens, tumours are complex ecosystems in which competition and cooperation between cells, combined with environmental pressures, shape the system dynamics \cite{Greaves2012,Marusyk2020} and its response to treatment. Poor treatment outcomes have indeed been related to tumour heterogeneity \cite{Dagogo-Jack2018}, with failure of therapies being associated with the presence and emergence of resistant phenotypes \cite{Baliu-Pique2020,Marusyk2020}. Resistant cells are indeed selected for during treatment and can drive subsequent relapse. Understanding the forces that drive such heterogeneity is therefore important for the design of more effective treatment strategies \cite{Baliu-Pique2020,Cajal2020,Marusyk2020}.

The concept of cancer stem cells (CSCs) was originally introduced to describe the hierarchical organization of tumours, motivated by analogies with normal tissue development \cite{Aponte2017,Greaves2012}. While differentiated cells (DCs) have a finite clonogenic capacity, (\emph{i.e.}, after dividing a finite number of times they exit the cell-cycle and become terminally differentiated (TDCs)), CSCs can give rise to cells with the same clonogenic capacity when they divide. This capacity for self-renewal makes CSCs key drivers of tumour initiation, progression and relapse. The coexistence of CSCs, DCs and TDCs in tumours is a source of intra-tumour heterogeneity. Furthermore, CSCs can contribute to heterogeneity because they can adapt to stressful environmental conditions (such as nutrient starvation or treatment) by changing their metabolism and/or entering a quiescent state \cite{Aponte2017,Fanelli2020,Sousa2019}. From this point of view, CSCs have the ability to survive standard treatment by transiently changing their behaviour \cite{Najafi2019} and can transmit these changes to their differentiated progeny resulting in a differentiated tumour bulk that is also resistant to treatment. 

Hypoxia, i.e., abnormally low oxygen levels, is commonly found in solid tumours due to aberrant vascularisation \cite{Hockel2001,Petrova2018}. As oxygen diffuses into a tumour and is consumed by cells, spatial gradients in oxygen levels are established and regions of chronic or transient hypoxia may emerge. The clinical implications of hypoxia and its link to several hallmarks of cancer \cite{Hanahan2000,Hockel2001,Ruan2009}, such evasion of apoptosis, genetic instabilities, induction of angiogenesis and metastasis, have long been known. The response of cells to hypoxia is in part regulated by \emph{hypoxia inducible factors} (HIFs) \cite{Ruan2009}, which can influence cancer cell metabolism, proliferation and ``stemness'' (a term that we use hereafter when referring to phenotypic variation of cancer cells). It has been shown that hypoxia facilitates the formation of stem-like cells by up-regulating stem-related genes in cancer cells in a HIF-dependent manner \cite{Axelson2005,Koh2011,Heddleston2010}. \emph{De novo} formation of CSCs can therefore occur in tumours in response to hypoxia. Indeed, hypoxia-mediated stress can drive cells to acquire a more aggressive, stem-like phenotype \cite{Pisco2015}. 

Mathematical models are a useful tool for studying the complex interplay between spatio-temporal variation in the tumour micro-environment and phenotypic heterogeneity and how this interplay impacts a tumours' growth dynamics and response to treatment. The critical role of space in shaping tumour composition and/or treatment outcome has been studied in previous works, where a wide range of mathematical formalisms has been used: from discrete agent based models \cite{Bacevic2017,Bull2020,Damaghi2021,Noble2019,Scott2014,Strobl2021} to continuum models of different degree of complexity \cite{Astanin2009,Fadai2020,Frieboes2010,Greenspan1972,Lewin2020,Villa2021}. In particular, phenotypic-structured models have been shown to be a useful tool for understanding intra-tumour phenotypic heterogeneity \cite{Ardaseva2020,Chisholm2016,Chisholm2015,Fiandaca2021,Hodgkinson2019,Lorenzi2016,Lorz2014,Shen2020,Stace2020,Villa2021}. These models comprise partial integro-differential equations in which changes in the tumour heterogeneity are mediated by proliferation, competition and epigenetic mutations. Such frameworks have been used to understand how oxygen gradients shape metabolic heterogeneity \cite{Fadai2020,Villa2021} as well as the consequences of resistance to chemotherapy \cite{Lorenzi2018,Villa2020}. 

In our previous work~\cite{CELORA2021}, we developed a stemness-structured model to investigate how heterogeneity in stemness levels affects the growth dynamics of tumours in a well-mixed environment. In our model, a cell's phenotypic state is described by a continuous variable, $s$, such that $s=0$ correspond to CSCs and $s=1$ to TDCs. Furthermore, cells may change their state in response to random and environmental-induced epigenetic changes, which are represented respectively via diffusion and advection along the $s$-axis. When local oxygen levels are sufficiently high, the advection flux favours differentiation; on the contrary, when oxygen levels are low the advection term favours de-differentiation towards a stem-like phenotype. The differential effects effects of radiotherapy (RT) are also included by assuming that cell radio-sensitivity depends on the stemness level of cells; in line with experimental evidence \cite{Arnold2020,Diehn2009,Galeaz2021}, CSCs are the most radio-resistant subpopulation. 
In this paper, we extend our earlier work to account for oxygen diffusion and consumption in the tumour and investigate how spatial variation in local oxygen levels affects phenotypic (\emph{i.e.},``stemness'') heterogeneity within a tumour. Using a combination of numerical techniques, we find that the interplay between physical and phenotypic space gives rise to more complex dynamics than those observed in our earlier work~\cite{CELORA2021}. In particular, features of the spatial model, such as multistability
and cusp bifurcations, disappear when spatial effects are neglected. We also investigate the combined effect of spatial and phenotypic heterogeneity on radiation outcome. We use numerical simulations to compare the efficacy of different treatment strategies and, in particular, their ability to prevent CSC-driven relapse. In doing so, we provide insight into the role that dose fractionation and scheduling may play in preventing relapse. 

The remainder of this paper is organised as follows. In \S\ref{sec:model} we present our stemness-structured spatially-resolved model and recast it in a partially non-dimensional form. In \S\ref{sec:numerics_no_radio}, we present typical numerical simulations of the model, which highlight the influence of the initial conditions in determining the long-time behaviour of the system in the absence of treatment. The observed multistability of the model is further investigated in \S\ref{sec:bifurcation}, where we numerically compute steady state solutions and use continuation techniques to identify how equilibria and their linear stability change as key model parameters vary. In \S\ref{sec:dynamics_with_radio}, we investigate tumour characterised by different levels of heterogeneity responds to standard radiotherapy protocols. how applying treatment can drive the eradication of the tumour by perturbing an initially stable state where the tumour persists. We find that the rate at which tumour cells consume oxygen plays a critical role in dictating the response to treatment, with high oxygen consumption rates, \emph{i.e.}, higher spatial-heterogenity of the oxygen levels in the tumour, correlating with poor outcome. In \S\ref{sec:conclusion}, we summarise our findings and outline future research directions.  The appendices offer a number of details
about the numerical methods, the parameter values and the bifurcation
analysis. 
%%%%%%%%%%%%%%%%%%%%%%%%%%%%%%%%%%%%%%%%%%
\section{Model Development}
\label{sec:model}
\begin{figure}[tb]
	\centering
	\includegraphics[width=0.95\textwidth]{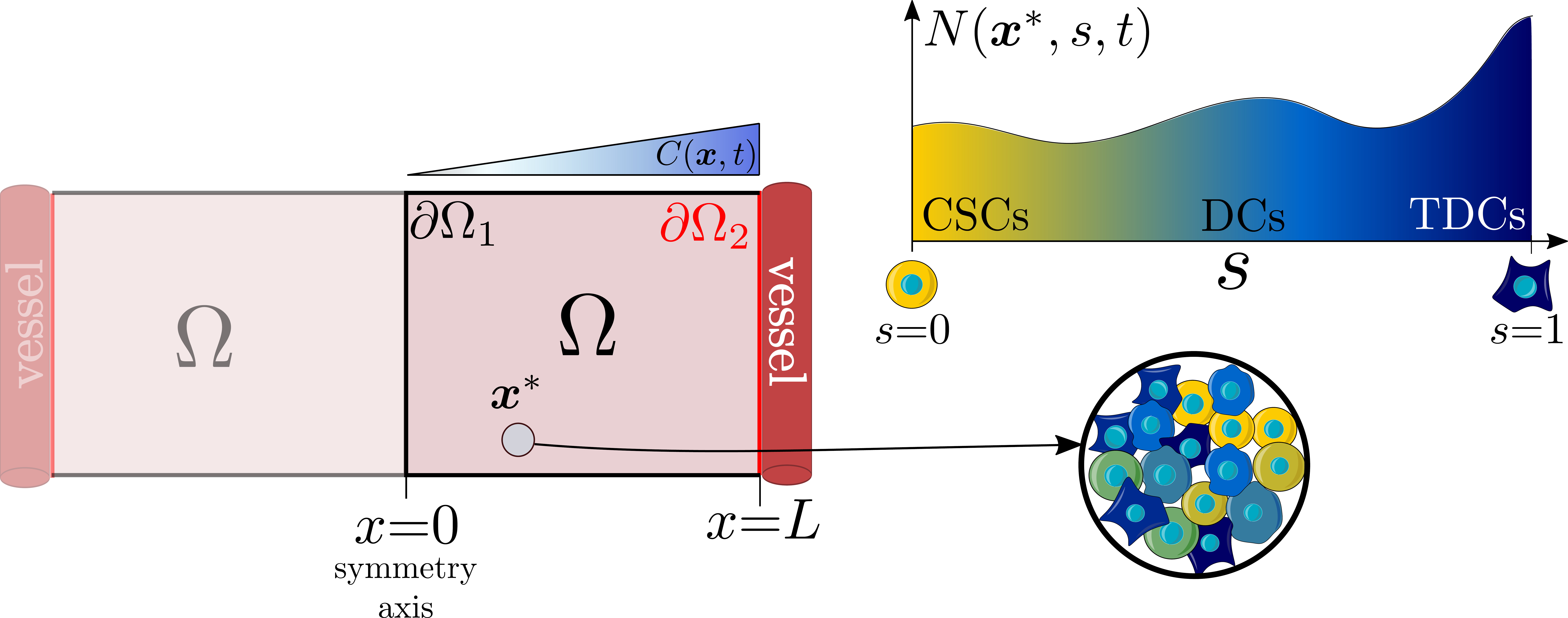}
	\caption{Schematic representation of the model structure: we consider a fixed slice of tissue delimited by two vessels at location $x=\pm L$. We reduce the model to a 1D Cartesian geometry by assuming that there is no dependency on the $y$ direction. Tumour cells living in this region are classified according to their stemness level $s$ where $s=0$ corresponds to fully stem-like cells while $s=1$ to cells that are terminally differentiated. At each location $x^*$ in space the density of cell with phenotype $s$ is denoted by $N(x^*,s,t)$ whose time evolution is described by the system~(\ref{eq:spatial_mod_no_scale}).}
	\label{fig:schematic}
\end{figure}
As shown in Fig.~\ref{fig:schematic}, we consider a fixed tissue region in which oxygen is supplied by blood vessels located on the domain boundaries. For simplicity, we consider a 1D Cartesian domain of length $2L$ and assume that the model variables depend on the spatial location $x\ [\mbox{mm}]$, where $x\in[-L,L]$, and time $t\, [hr]$. 
The state of a tumour cell in phenotypic space is described by its stemness level $s$ (we view $s$ as a dimensionless independent variable); stemness levels influence a cell proliferation capacity, resistance to treatment (here radiotherapy) and response to physiological stresses (here \emph{hypoxia}). The local cell phenotypic distribution $N=N(x,s,t)\ [\mbox{cells} \mbox{ mm}^{-1}]$ indicates the number of tumour cells with phenotype $s$, at spatial location $x$ and time $t$. We further introduce $\Phi=\Phi(x,t)\ [\mbox{cells} \mbox{ mm}^{-1}]$, which represents 
the total number of tumour cells at location $x$ and time $t$:
\begin{equation}
\Phi(x,t)=\int_0^1 N(x,s,t) \, ds.
\end{equation}
The evolution of $N$ is dictated by cell proliferation and death 
as well as spatial ($J_x=J_x(x,s,t)$) and phenotypic ($J_s=J_s(x,s,t)$) fluxes along the $x$ and $s$ axis, respectively. Cell behaviour is mediated by local levels of oxygen $C=C(x,t)\ [mmHg]$. As oxygen diffuses into the tissue, it is consumed by cells at a rate that may depend on their phenotype, causing different oxygen environments, or \emph{niches}, to form. Under these assumptions, we have that the time evolution of $N$ and $C$ is determined by the following system of coupled non-linear and non-local parabolic partial differential equations (PDEs):
\begin{subequations}
	\begin{align}	
	\frac{\partial C}{\partial t} = D_{xc}\frac{\partial^2 C}{\partial x^2}-\underbrace{\int_0^1 \Gamma(s,C) N(x,s,t) \, ds}_{\mbox{net oxygen consumption}},\label{eq:Cdim}\\
	\frac{\partial N}{\partial t}+ \frac{\partial J_x}{\partial x}+ \frac{\partial J_s}{\partial s}= F(s,C,\Phi,t)  N.\label{eq:Ndim}
	\end{align}
	In Eq.~(\ref{eq:Cdim}), the positive constant $D_{xc}\ [\mbox{mm}^2\mbox{hr}^{-1}]$ is the diffusion coefficient of oxygen and the function $\Gamma=\Gamma(s,C)\ [\mbox{mmHg}(\mbox{cell}\mbox{ hr})^{-1}]$ represents the rate at which cells of phenotype $s$ exposed to local oxygen levels $C$ consume oxygen. Focusing now on Eq.~(\ref{eq:Ndim}), and following \cite{Fiandaca2021,Villa2021}, we assume that cells simply move randomly in space, in which case the spatial flux $J_x$ is given by:
	\begin{align}
	J_x= - D_{xn} \frac{\partial N}{\partial x},
	\end{align}
	where the positive constant $D_{xn}\ [\mbox{mm}^2\mbox{hr}^{-1}]$ represents the diffusion coefficient of cells. What distinguishes our model from previous works is the inclusion of the phenotypic flux $J_s$, which comprises two effects: random epigenetic mutations and environment-mediated changes in cell stemness. These are represented, respectively, via diffusion and advection terms so that $J_s$ is given by:
	\begin{align} 
	J_s= - D_{sn} \frac{\partial N}{\partial s}+ V_s(s,C)N.
	\end{align}
	Here, the positive constant $D_{sn}\ [\mbox{hr}^{-1}]$ represents the rate at which random epigenetic mutations take place, while the velocity $V_s=V_s(s,C)\ [\mbox{hr}^{-1}]$ captures the rate at which cells of phenotype $s$ alter their stemness level in response to local oxygen levels. Focusing on the right-hand side of Eq.~(\ref{eq:Ndim}), the function $F=F(s,C,\Phi,t)\ [\mbox{hr}^{-1}]$ describes the net rate of cell proliferation. As such, it also captures the effect of radiotherapy, which we assume to be administered in $M$ doses $d_i\ [Gy]$ at discrete times $t_i$ with $i=1,\ldots,M$. Under these assumptions, we define $F$ as:
	\begin{align}
	\begin{aligned}
	F(s,C,\Phi,t)= \underbrace{P(s,C)\left(1-\frac{\Phi}{\Phi_{max}}\right)}_{\mbox{proliferation}} &-\underbrace{K(s,C)}_{\mbox{cell death}} \\&-\underbrace{\sum^{M}_{i=1} \log\left(\frac{1}{S_{RT}(s,C,\Phi;d_i)}\right)\delta(t-t_i)}_{\mbox{radiotherapy}}.\label{eq:F}
	\end{aligned}
	\end{align}
	In Eq.~(\ref{eq:F}), $\Phi_{max}\ [\mbox{cells}/\mbox{mm}]$ represents the local carrying capacity of the tissue, so that if the initial condition is such that $\Phi(x,0) \in [0,\Phi_{max}]$ then $\Phi(x,t) \in [0,\Phi_{max}]$ for all $t>0$. The associated logistic-growth term captures growth inhibition due to competition for space. Locally, cells proliferate at a rate $P=P(s,C)\ [hr^{-1}]$ and die at a rate $K=K(s,C)\ [hr^{-1}]$ both rates being modulated by the stemness levels $s$ and the local oxygen levels, $C$. The final sink term in Eq.~(\ref{eq:F}) is associated with cell death due to radiotherapy (RT); here $S_{RT}=S_{RT}(s,C,\Phi;d_i)\, \in[0,1]$ denotes the fraction of cells that survive exposure to a dose $d_i$ of RT. As discussed in \cite{CELORA2021}, we use the term $\log(1/S_{RT})$ although other forms have been proposed in the literature. For example, in \cite{Lewin_2020}, the term $(1-S_{RT})$ is used. The improved accuracy of the logarithmic functional form was argued in~\cite{CELORA2021}. We postpone the specification of functional forms in the model (\emph{i.e.}, $V$,$P$, $K$, and $SF$) until we have rescaled the model equations. 
	
	In line with the geometry presented in Fig.~\ref{fig:schematic}, we impose the following boundary conditions:
	\begin{align}
	\left.D_{sn} \frac{\partial N}{\partial s}-N V_s\right|_{s\in\left\{0,1\right\}} = 0,& \quad x\in [-L,L],\, t>0,\label{BC1}\\
	\left.\frac{\partial N}{\partial x}\right|_{x=\pm L}=0,& \quad s\in(0,1),\, t>0,\label{BC2}\\[2pt]
	C(\pm L,t)=C_{\infty},& \quad t>0. \label{BC3}
	\end{align}% 
	Eqs.~(\ref{BC1})-(\ref{BC2}) state that the phenotypic and spatial fluxes of the cells vanish on the relevant domain boundaries. In Eq.~(\ref{BC3}), the supply of oxygen (from vessels located at tissue boundaries) is modelled by prescribing a constant oxygen concentration, $C_{\infty}$, on $x=\pm L$.
	Finally, we close the governing equations by imposing the following initial conditions:
	\begin{align}
	N(x,s,0)= N_0(x,s)& \quad x\in[-L,L],\, s\in(0,1),\\[2mm]
	C(x,0)= C_0(x)&\quad  x\in[-L,L].
	\end{align}\label{eq:spatial_mod_no_scale}
\end{subequations}
\paragraph{Non-dimensional model formulation}
We now recast the model in a partially dimensionless form; all variables are rescaled except for time $t$ which is measured in hours. We continue to view time as a dimensional variable to facilitate interpretation of model simulations, particularly when RT is included. We choose the following scalings for the other model variables
\begin{equation}
X=\frac{x}{L},\quad n=\frac{N}{\Phi_{max}}, \quad \phi=\frac{\Phi}{\Phi_{max}}, \quad c=\frac{C}{C_\infty}. \label{eq:scaling}
\end{equation}
Substituting~(\ref{eq:scaling}) into~(\ref{eq:spatial_mod_no_scale}) and exploiting the assumed symmetry about the axis $X=0$, we deduce that the evolution of the re-scaled cell phenotypic distribution $n$ is governed by
\begin{subequations}
	\begin{align}
	\frac{\partial n}{\partial t}= \parD{xn} \frac{\partial^2 n}{\partial X^2}+ \frac{\partial }{\partial s} \left(\parD{sn} \frac{\partial n}{\partial s}-v_s(s,c)n\right)+ \mathcal{F}(s,c,\phi,t) n,& \quad X\in (0,1),\  s\in(0,1),\  t>0,\label{eq:final_n}\\	\left.\parD{sn} \frac{\partial n}{\partial s}-n v_s\right|_{s\in\left\{0,1\right\}} = 0,& \quad X\in [0,1],\, t>0,&\label{bc:on_s}\\
	\left.\frac{\partial n}{\partial X}\right|_{X\in\left\{0,1\right\}}=0,& \quad s\in(0,1),\, t>0,\\[4pt]
	n(X,s,0)= n_0(X,s),& \quad X\in[0,1],\, s\in(0,1),&
	\end{align}
	where
	\begin{align}
	\phi(X,t)&=\int_0^1 n(X,s,t) \, ds,\\
	\mathcal{F}(s,c,\phi,t)&=
	\mathcal{P}(s,c)\left(1-\phi\right)-\mathcal{K}(s,c) 	-\sum^{M}_{i=1} \log\left(\frac{1}{\mathcal{S}_{RT}(s,c,\phi;d_i)}\right)\delta(t-t_i),\label{eq:F_adim}
	\end{align}\label{eq:rescaled_sys}
\end{subequations}
and the evolution of the normalised oxygen concentration $c$ is dictated by 
\begin{subequations}	
	\begin{align}
	\frac{\partial c}{\partial t} = \parD{xc}\frac{\partial^2 c}{\partial X^2}-\int_0^1 \gamma(s,c) n(X,s,t) \, ds,& \quad X\in (0,1),\  t>0,\\	
	c(1,t)=1, \quad \frac{\partial c}{\partial X}(0,t)=0,&\quad t>0,\label{bc:ox}\\[4pt]
	c(x,0)= c_0(x),& \quad  X\in[0,1].
	\end{align}\label{eq:rescaled_sysOx1}%
\end{subequations}
In Eqs.~(\ref{eq:rescaled_sys})-(\ref{eq:rescaled_sysOx1}), we have introduced the following (partially dimensionless) parameter groupings and functional forms:
\begin{equation}
\begin{aligned}
\parD{xn}&= \frac{D_{xn}}{L^2}, \quad  \parD{xc}= \frac{D_{xc}}{L^2}, \quad  \gamma(s,c)=\frac{\Gamma(s,c\,C_\infty)\Phi_{max}}{C_{\infty}},\\[2pt]
v_s(s,c)&=V_s(s,c\,C_\infty),\quad \mathcal{P}(s,c)=P(s,c\,C_\infty),\quad \mathcal{K}(s,c)=K(s,c\,C_\infty),\\[4pt]
\mathcal{F}(s,c,\phi,t)&=F(s,c\,C_{\infty},\phi\,\Phi_{max},t),\qquad  \mathcal{S}_{RT}(s,c,\phi;d_i)=S_{RT}(s,c\,C_\infty,\phi\,\Phi_{max};d_i).
\end{aligned}
\end{equation}

\subsection{Specification of functional forms.}
\label{subsec:FuncForm}
We now define the functional forms for $v_s$, $\mathcal{P}$, $\mathcal{K}$, $\gamma$ and $\PARA{SF}$. Some are similar to those used in \cite{CELORA2021} where we have studied the spatially well-mixed counterpart of Eqs.~(\ref{eq:rescaled_sys})-(\ref{eq:rescaled_sysOx1}). Therefore we refer to \cite{CELORA2021} for a more detailed discussion and here focus on the key differences. 

\begin{figure}
	\centering
	\includegraphics[width=0.7\textwidth]{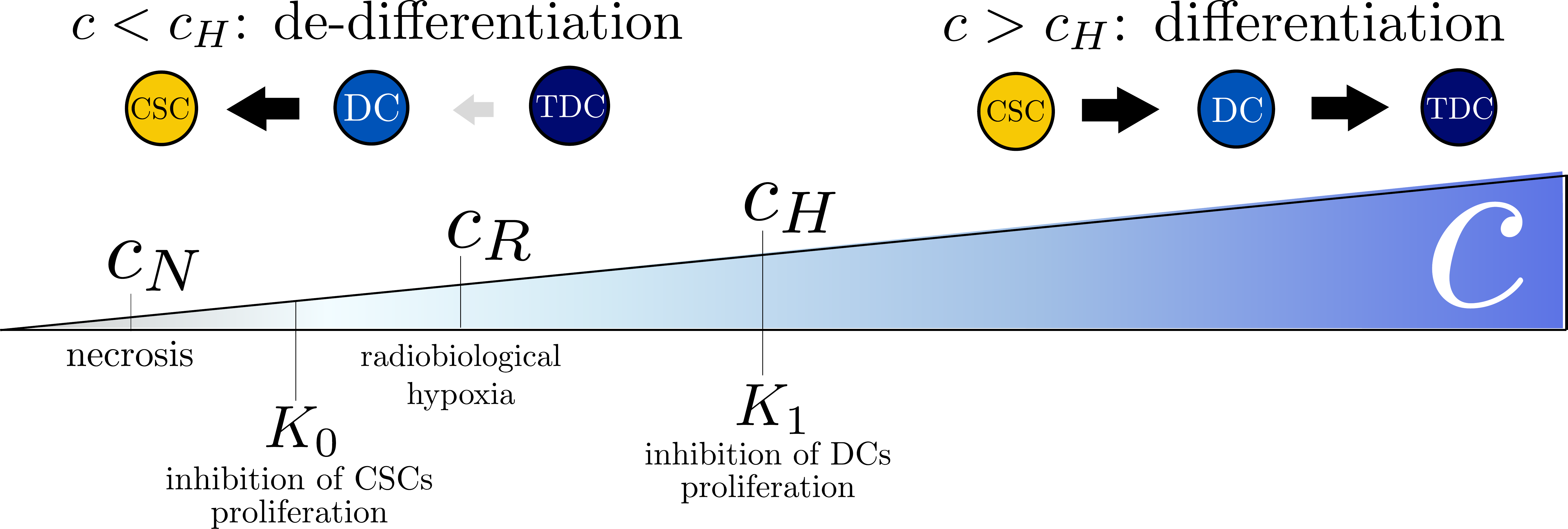}
	\caption{Schematic representing the role of oxygen in the model. Local oxygen levels mediate proliferation as well as death, the differentiation of cells and their sensitivity to radiotherapy. Note that the scaling of oxygen levels is here for illustrative purposes. Values associated with the thresholds can be found in Table~\ref{tab:parameters2}. }
	\label{fig:oxygen}
\end{figure}

To facilitate understanding of our model, we summarise in Fig.~\ref{fig:oxygen} the ways in which we assume local oxygen levels influence tumour cell behaviours. This, in turn, leads us to subdivide the tumour into four \emph{niches} according to local oxygen levels: normoxia ($c_H<c<c_\infty$), hypoxia ($c_N<c<c_H$), radio-biological hypoxia ($c_N<c<c_R$) and necrosis ($0\leq c<c_N$). 

\begin{figure}
	\begin{subfigure}{0.49\textwidth}
		\centering
		\includegraphics[width=0.95\textwidth]{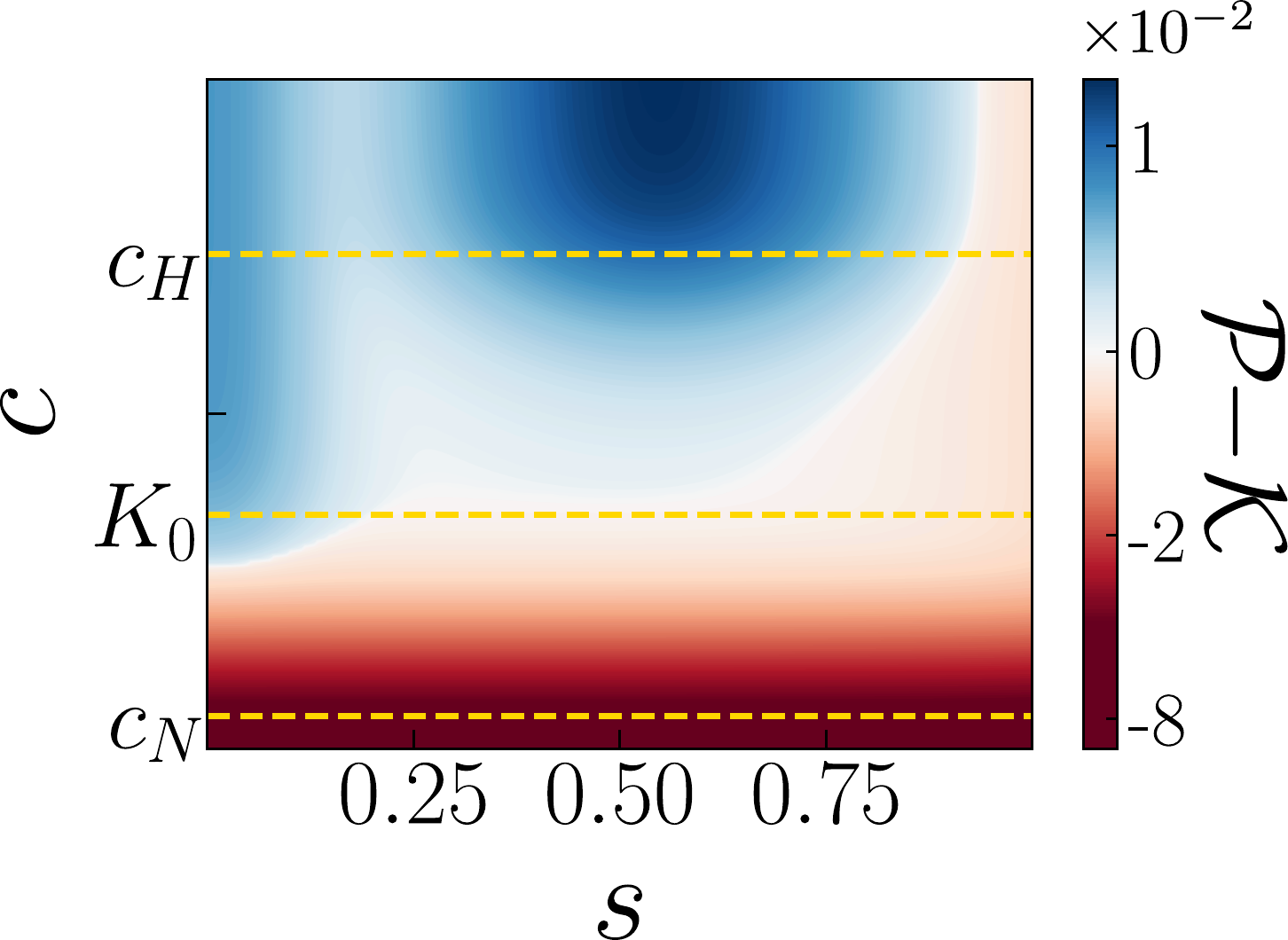}
		\caption{}
		\label{fig:netrate}
	\end{subfigure}
	\begin{subfigure}{0.49\textwidth}
		\centering
		\includegraphics[width=0.95\textwidth]{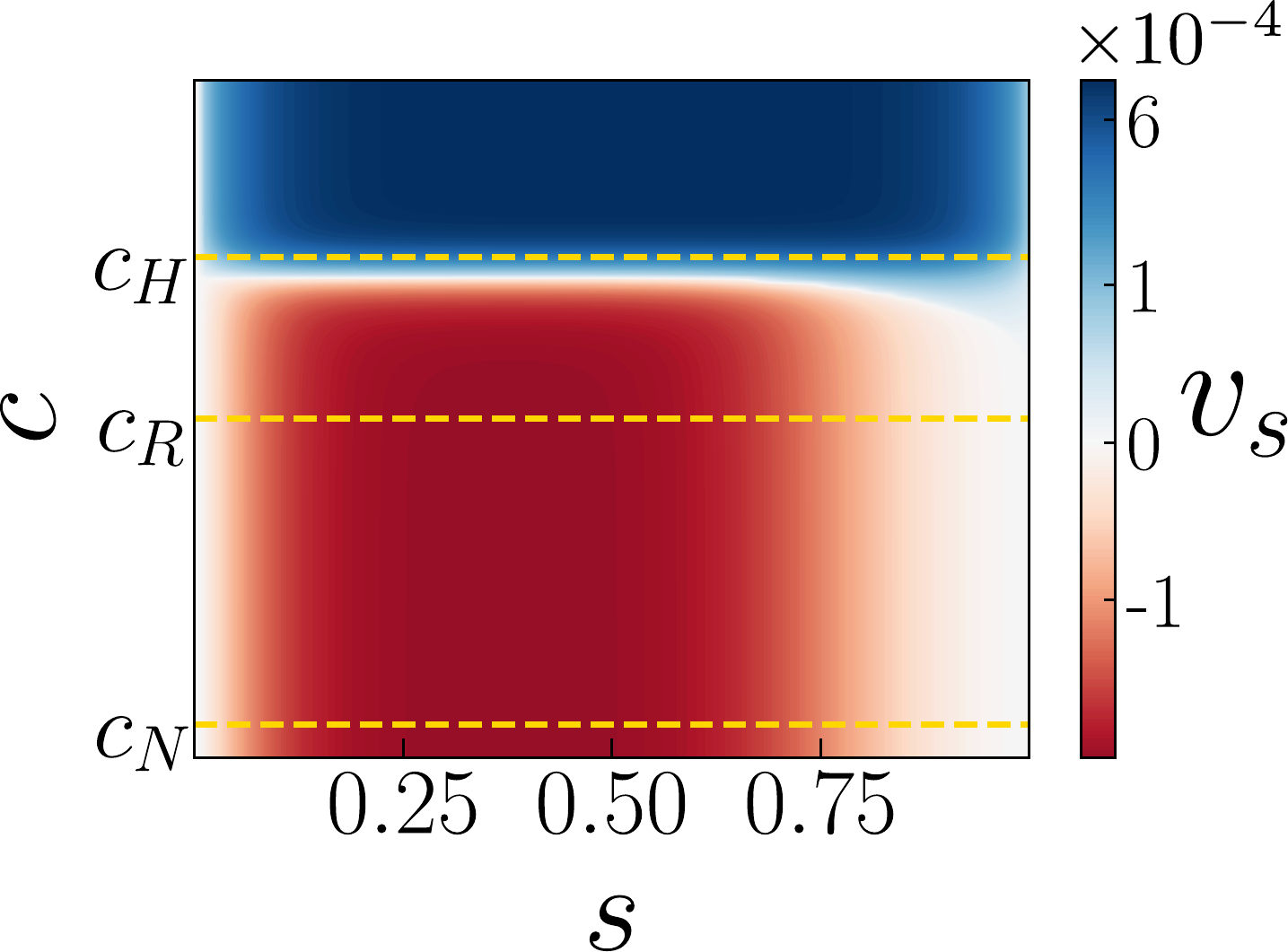}
		\caption{}
		\label{fig:vel}
	\end{subfigure}
	\caption{Surface plot of the functional (a) $\mathcal{P}(s,c)-\mathcal{K}(s,c)$ and (b) $v_s(s,c)$ with indication of the oxygen threshold that dictates their behaviour. Note that we use a logarithmic scale for the oxygen $c$. For the parameters used in the figure see Table~\ref{tab:parameters2}.}
	\label{fig:velprol}
\end{figure}

We define the proliferation rate $\PARA{P}$ and natural death rate $\PARA{K}$ in Eq.~(\ref{eq:F_adim}) as follows:  
\begin{subequations}
	\begin{align}
	\mathcal{P}(s,c)=p_0(c)\exp\left[-\frac{s^2}{g_0}\right]+p_1(c)\exp\left[-\frac{(s-s_{max})^2}{g_1}\right],\label{eq:P}\\[4pt]
	\mathcal{K}(s,c)= \underbrace{d_f \,e^{-k_f(1-s)}}_{\footnotesize\substack{ \mbox{natural cell}\\[2pt] \mbox{death}}} + \underbrace{d_n H_{\epsilon_{K}}(c_N-c)}_{\footnotesize\mbox{necrosis}}\label{eq:K},\\
	\mbox{where }p_i(c)=p_i^{max}\myfrac[2pt]{c^4}{K_{i}^4+c^4}, \quad i\in\left\{0,1\right\}.\label{eq:p_coeff_with_ox}\\
	\end{align}
	In Eq.~(\ref{eq:P}), a bimodal function describes the proliferation rate $\mathcal{P}$; the two peaks correspond to CSCs ($s=0$) and differentiated cells (DCs) ($s_{max}=0.55$) which proliferate at rates $p_0(c)$ and $p_1(c)$ respectively. We assume further that both $p_0$ and $p_1$ are increasing, saturating functions of $c$ (see Eq.~(\ref{eq:p_coeff_with_ox})). In physiological oxygen conditions ($c\approx 1$), CSCs and DCs proliferate at their maximum rates, $p^{max}_0$ and $p^{max}_1$, respectively. The parameters $K_0$ and $K_1$ represent the oxygen concentrations at which CSCs and DCs proliferate at half their maximum rates (see Fig.~\ref{fig:oxygen}). We fix $p^{max}_0<p^{max}_1$ and $K_0<K_1$ so that DCs are more proliferative than CSCs in normoxia, in line with \cite{Dhawan2016,Hoffmann2008}, but their growth is more significantly inhibited in hypoxia ($c\lessapprox K_1\equiv c_H$). In contrast, CSCs proliferate more slowly than DCs in normoxia, but they are less sensitive hypoxia (since $K_0<K_1$) and continue to proliferate albeit at a low rate. In this way, we capture the higher plasticity of CSCs and, in particular, their ability to adapt to hypoxia \cite{Chae2018,Garnier2019,Snyder2018}. In Eq.~(\ref{eq:K}), we account for two mechanisms of cell death: natural cell death and necrosis. While natural cell death is assumed to be independent of oxygen levels $c$ and only to affect TDCs ($s\approx 1$), necrosis impact all cells regardless of their stemness level $s$ only when oxygen levels are extremely low ($c\lessapprox c_N\ll 1$). 
	To capture the effect of necrosis we introduce the continuous step function, $H_{\epsilon}$:
\end{subequations}
\begin{equation}
H_\epsilon(x)= \frac{1}{2}+\frac{1}{2}\tanh\left(x\epsilon^{-1}\right),\label{eq:H}
\end{equation}
where the parameter $\epsilon$ determines how rapidly $H_{\epsilon}$ switches between its extremal values ($H_{\epsilon}(x)\approx 1$ for $x>\epsilon$ and $H_{\epsilon}(x)\approx 0$ for $x<-\epsilon$). In Eq.~(\ref{eq:K}), we assume that the transition between the hypoxic and necrotic niches is rapid by setting $\epsilon_{K}=0.01$. In Fig.~\ref{fig:velprol}(a), we plot the net proliferation rate, in the absence of competition (\emph{i.e.}, $\mathcal{P}-\mathcal{K}$), and show how it depends on the local oxygen levels ($c$) and phenotype ($s$). 

Following \cite{CELORA2021}, we propose the following functional form for the phenotypic advection velocity $v_s$: 
\begin{subequations}
	\begin{align}
	v_s(s,c)=v^+_s(s) H_{\epsilon_v}(c-c_H)-v^-_s(s)H_{\epsilon_v}(c_H-c),\\[3pt]	
	v^+_s(s)= \myfrac[2pt]{V_+}{V^*_+} \tanh\left(\myfrac[3pt]{s^{\omega_+}}{\xi_+}\right)\tanh\left(\myfrac[2pt]{1-s}{\xi_+}\right),\label{eq_ad_norm}\\[3pt]
	v^-_s(s)=\myfrac[2pt]{V_-}{V^*_-} \tanh\left(\myfrac[2pt]{s}{\xi_-}\right)\tanh\left(\myfrac[3pt]{(1-s)^{\omega_-}}{\xi_-}\right),\label{eq_ad_hyp}
	\end{align}\label{eq_ad}%
\end{subequations}%
In Eqs.~(\ref{eq_ad_norm})-(\ref{eq_ad_hyp}), the normalizing factors $V_\pm^*$ are defined so that $\max\limits_{s\in[0,1]}(v_s^\pm)=V_\pm$, where $V_\pm$ correspond to the maximum absolute values of the velocities with which cells differentiate ($V_+$) and de-differentiate ($V_-$). The positive constants $\xi_\pm$ determine how rapidly the velocity profiles switch near $s=0$ and $s=1$. 
The exponents $\omega_\pm$ are positive integers: $\omega_+=1$ and $\omega_-=2$ (this choice is based on the analysis performed in \cite{CELORA2021}). As in \cite{Picco2017}, we assume that there is a threshold oxygen concentration $c_H$ that separates the tumour into two \emph{niches}. 
As illustrated in Fig.~\ref{fig:velprol}(b), when $c>c_H$ cells differentiate and progress towards $s=1$ (i.e., $v_s>0$); when $c<c_H$, cells instead de-differentiate and assume stem-like traits (i.e., $v_s<0$). There is a smooth transition between the two niches, capture by the function $H_{\epsilon_v}$, which is defined as in Eq.~(\ref{eq:H}), with $\epsilon_v=0.05$. 

To complete the model formulation for tumour growth in the absence of treatment, we must specify the functional form of the oxygen consumption rate $\gamma=\gamma(s,c)$. As already mentioned, in practice oxygen consumption might depend on the stemness variable $s$. However, in the absence of suitable experimental data to define this dependence, we follow~\cite{Scott2014} and assume $\gamma=\gamma(c)$.
Furthermore, following \cite{Greenspan1972,Lewin2020}, we suppose that $\gamma$ drops rapidly when oxygen levels fall below the necrotic threshold $c=c_N$:
\begin{equation}
\gamma=\gamma_0 H_{\epsilon_\gamma}(c-c_N).\label{eq:def_gamma}
\end{equation}
In Eq.~(\ref{eq:def_gamma}), $\gamma_0$ is the maximum oxygen consumption rate for cells and $\epsilon_\gamma=0.05$. We rewrite Eq.~(\ref{eq:def_gamma}) in terms of the diffusion coefficient $\parD{xc}^{-1}$ by introducing the non-dimensional parameter $\gamma^*=\gamma_0\parD{xc}^{-1}$. The latter gives the ratio between the timescales for oxygen diffusion and oxygen consumption. As we will see in \S\ref{sec:bifurcation}, this parameter plays a key role in determining the long time behaviour of the system.

\paragraph{The survival fraction}
Sensitivity of cells to radiotherapy is modulated by environmental conditions \cite{Sorensen2020,Suwa2021} as well as cell-intrinsic factors \cite{Arnold2020,Galeaz2021,Syljuasen2019,Tang2018}. We start by considering the standard linear-quadratic model which states that the fraction of cells that survive a single dose $d$ of radiation is given by \cite{McMahon2018}:
\begin{subequations}
	\begin{align}
	\ln \left(\PARA{S}_{RT}(c,s,\phi;d)\right)= - \alpha(c,s,\phi) d - \beta(c,s,\phi) d^2.\label{eq:LQ}
	\end{align}%
	In what follows, we suppose that $\alpha [\Gray^{-1}]$ and $\beta [\Gray^{-2}]$ are dependent on the local environment, and, in particular, the oxygen concentration $c$, the cell density $\phi$, and the stemness of the cell, $s$. Consequently, in our model the overall tumour radio-sensitivity depends on the tissue composition and the environment, both of which are themselves affected by treatment. 
	
	We decompose $\alpha$ (and, analogously, $\beta$) into three factors, which account for different mechanism of radio-resistance:
	\begin{align}
	\alpha(s,c,\phi)=\alpha_1(c)\alpha_2(s)\alpha_3(c,s,\phi), \quad \beta(s,c,\phi)=\beta_1(c)\beta_2(s)\beta_3(c,s,\phi).
	\end{align}
	First, it is well known that low oxygen levels ($c<c_R$) can adversely impact radiotherapy efficacy; oxygen is an essential reactant for the fixation of radio-induced DNA damage and subsequent cell death \cite{Wenzl2011,McMahon2018,Sorensen2020,West2019}. This effect is commonly incorporated into the survival fraction by introducing the \emph{oxygen enhancement ratio} ($\mbox{OER}$)\cite{Wenzl2011,Lewin_2020}. The OER is a constant parameter whose value usually ranges between $1$ and $4$; it is defined to be the ratio of the radiation dose that needs to be given in anoxia compared to normoxia to achieve the same biological effects (i.e., the same overall survival fraction) \cite{Wenzl2011}. As in \cite{Lewin_2020}, we assume that the OER plays a role only in regions of \emph{radio-biological hypoxia} (i.e., regions where $c\leq c_R$), and propose the following functional form for $\alpha_1(c)$ and $\beta_1(c)$:
	\begin{align}
	\alpha_1(c)= \begin{cases}1,& \quad c>c_R\\
	\mbox{OER}^{-1},&\quad c\leq c_R\end{cases},\quad \beta_1(c)= \begin{cases}1,& \quad c>c_R\\
	\mbox{OER}^{-2},&\quad c\leq c_R\end{cases}
	\end{align}
	where the threshold $c_R$ is chosen such that $c_N<c_R<c_H$ (see Fig.~\ref{fig:oxygen}).
	As in \cite{CELORA2021}, we account for the intrinsic radio-resistance of CSCs \cite{Arnold2020,Diehn2009,Galeaz2021,Olivares-Urbano2020} by proposing the following functional forms for $\alpha_2(s)$ and $\beta_2(s)$:
	\begin{align}
	\alpha_2(s)= \alpha_-+\Delta\alpha \tanh(\kappa s),\quad \beta_2(s)= \beta_-+\Delta\beta \tanh(\kappa s), \label{eq:al_bet2}
	\end{align}%
	In Eq.~(\ref{eq:al_bet2}), the positive constants $\alpha_-$ and $\beta_-$ represent the radio-sensitivity of CSCs (\emph{i.e.}, $s\approx 0$) while $\Delta\alpha$ and $\Delta\beta$ denote the difference in radio-sensitivity between CSCs and more differentiated cell phenotypes ($s>0.5$). 
	In contrast to \cite{CELORA2021}, we include an additional term to account for the increased sensitivity of rapidly proliferating cells compared to growth-arrested cells~\cite{Barendsen2001,Enderling2009,Franken2013} by assuming 
	that $\alpha_3(c,s,\phi)$ and $\beta_3(c,s,\phi)$ depend on the effective cell proliferation rate in the following way:
	\begin{align}
	\alpha_3(c,s,\phi)=\beta_3(c,s,\phi)=\left(1+\frac{\mathcal{P}(c,s)(1-\phi)}{\mathcal{P}_{max}}\right).\label{eq:SF3}
	\end{align}
	In Eq.~(\ref{eq:SF3}), $\PARA{P}_{max}=\max_{c,s} \PARA{P}=p_0^{max}$. Equation~(\ref{eq:SF3}) implies that rapidly proliferating cells are more sensitive to radiation than cells which proliferate less rapidly. 
	
	The combined effect of the three radio-resistance mechanisms is illustrated in Fig.~\ref{fig:surv_frac}. Note that we show plots of $\alpha$ only since $\beta$ has a qualitatively similar effect. When oxygen levels drop below $c_R$ (see lighter line in Fig~\ref{fig:surv_frac}), the radio-sensitivity of the cells also falls. For milder levels of hypoxia (e.g., $c_R<c<c_H$), cell proliferation is inhibited for all phenotypes $s$ so that the dominant mechanism modulating radio-sensitivity is the intrinsic radio-sensitivity of the cells (\emph{i.e.}, $\alpha\sim \alpha_2(s)$). Spatial competition is relevant at higher oxygen concentration (e.g, $c>c_H$). Where spatial competition is low ($\phi\approx 0$), differentiated cells ($s\approx 0.5$) are highly radio-sensitive; by contrast, TDCs are less sensitive as they are not proliferating. However, as the population 
	grows, competition for space inhibits cell proliferation so that TDCs and differentiated cells have similar responses to radiotherapy. 
	\begin{figure}[ht]
		\centering
		\includegraphics[width=\textwidth]{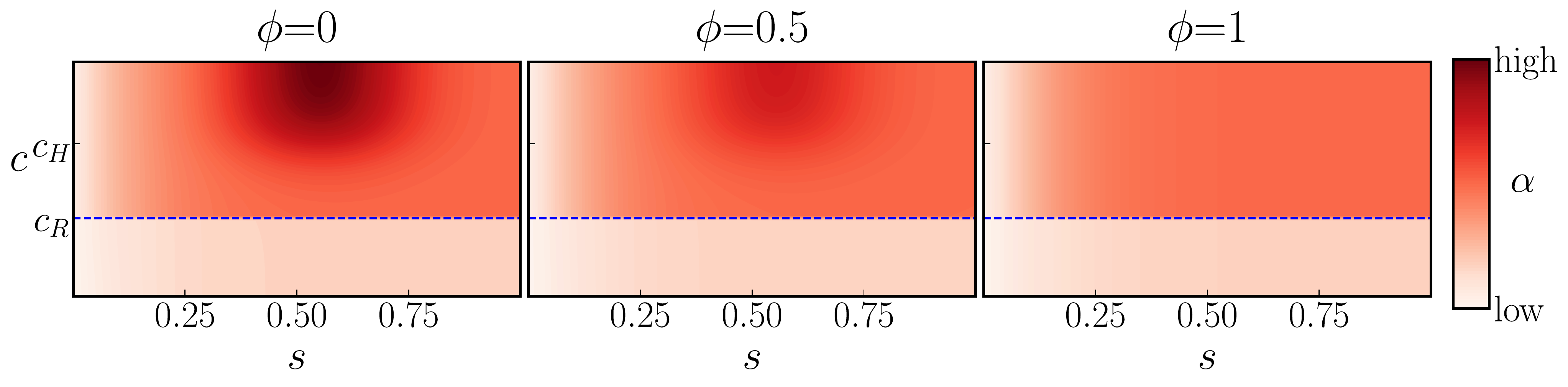}
		\caption{Series of plot showing how, for fixed value of $\phi$, $\alpha=\alpha(s,c,\phi)$ (defined by Eq.~(\ref{eq:LQ})) changes as the phenotypic variable $s$ and the oxygen concentration $c$ vary. We plot $\alpha$ for different values of the re-scaled cell density $\phi$ to show also how competition for space modulates the radio-sensitivity.  Parameter values set as in Tables~\ref{tab:parameters1}-\ref{tab:parameters2}.}
		\label{fig:surv_frac}
	\end{figure}
\end{subequations}

\subsection{Numerical methods and interpretation of the numerical simulations}

When performing dynamic simulations of Eqs~(\ref{eq:rescaled_sys})-(\ref{eq:rescaled_sysOx1}), most parameters are fixed to the values listed in Tables~\ref{tab:parameters1}-\ref{tab:parameters2}; where possible these values have been taken from the literature.

Based on standard values reported in the literature and considering a tissue region that is several micrometers wide ($L\approx 200\, \mu m$), we find that the time-scale of oxygen diffusion in the tumour region ($\parD{xc}^{-1}=L^2 D_{xc}^{-1}$) is of the order of seconds or minutes, which is much faster than the time-scale of cell proliferation ($\mathcal{P}^{-1}$ of the order of hours and days) which is the timescale of interest here. We exploit this separation of timescales to simplify the equations governing the oxygen dynamics. As is standard in mathematical models of tumour growth \cite{Bull2020,Lewin2020}, we assume that the oxygen distribution is in quasi-equilibrium so that $c$ satisfies
\begin{equation}
\frac{\partial^2 c}{\partial X^2}-\gamma^* \phi(X,t)H(c-c_N)=0, \quad c(1,t)=0, \quad \frac{\partial c}{\partial X}=0.\label{eq:c_stationary}
\end{equation}%
Under the quasi-steady-state approximation, the oxygen quickly adapts is spatial distribution to changes in the tumour composition. As a result, we do not need to impose initial conditions for $c$. Under this assumption, the oxygen profile is primarily characterised by the non-dimensional parameter $\gamma^*=\gamma_0L^2 D_{xc}^{-1}$, which we recall measures the ratio of the time scales for oxygen diffusion in a tissue of size $L$ to the time scale for oxygen consumption ($\gamma_0^{-1}$). 
Since we are interested in understanding how spatial variation in oxygen levels affects the tumour's phenotypic composition, we view $\gamma^*$ as a free parameter, which varies between different tumour cell populations. While the oxygen diffusion coefficient is typically considered as a constant $D_{xc}$ (see Table~\ref{tab:parameters1}), oxygen consumption rates ($\gamma_0$) are likely to be tumour- and patient-specific. For this reason, when comparing different values of $\gamma^*$, we will implicitly assume that the tumours have the same `typical' length scale, $L$, and oxygen diffusion coefficient but differ in their oxygen consumption rates $\gamma_0$.

Unless otherwise stated, we initialise the simulations by assuming that, at $t=0$, the cells are distributed uniformly in space and normally along the stemness axis, with mean stemness $s_0\in(0,1)$ and variance $\sigma_s>0$. Thus, we prescribe
\begin{equation}
n(s,X,t=0) = n_0(s,X)= \frac{M_0}{\sqrt{2\pi}\sigma_s}\exp\left[-\frac{(s-s_0)^2}{2\sigma^2_s}\right].\label{eq:initial}
\end{equation}
where $M_0>0$ is the initial tumour burden, while $s_0$ and $\sigma_s^2$ are chosen so that the mass of the Gaussian profile lies inside the stemness domain $s \in (0,1)$ and the Neumann boundary conditions~(\ref{bc:on_s}) are satisfied, up to a negligible (exponentially small) correction.

We use the method of lines to solve Eqs.~(\ref{eq:rescaled_sys})-(\ref{eq:c_stationary}). In more detail, we first discretise Eqs.~(\ref{eq:rescaled_sys}) and~(\ref{eq:c_stationary}) using finite volume and central finite differences to discretise the phenotypic ($s$) and spatial ($X$) axis, respectively. We then advance Eq.~(\ref{eq:final_n}) in time by using an implicit multi-step time-stepping method based on backward differentiation formulae (BDF) implemented in the \emph{scipy} python package, which is suitable for stiff equations. At each time point $t$, given the discrete form of the solution $n(s,X,t)$, we can compute $\phi(X,t)$. We use the latter to solve for the stationary distribution of the oxygen concentration $c$ via a fixed point iteration method. A detailed description of the numerical method used can be found in~\ref{App:numerics}. 

\paragraph{Interpretation of the numerical results}
The model has several independent variables which can make it difficult to simultaneously visualise the evolution of the tumour's spatial and phenotypic distributions. 

As a measure of tumour load, we introduce the total tumour burden $M=M(t)$:
\begin{equation}
M(t)=\int_0^1\int_0^1 n(X,s,t) dXds.\label{eq:tumourburden}
\end{equation} 
We note that $M\Phi_{max}L$ represents the total number of tumour cells in the slice of tissue.

\begin{figure}[ht]
	\centering
	\includegraphics[width=\textwidth]{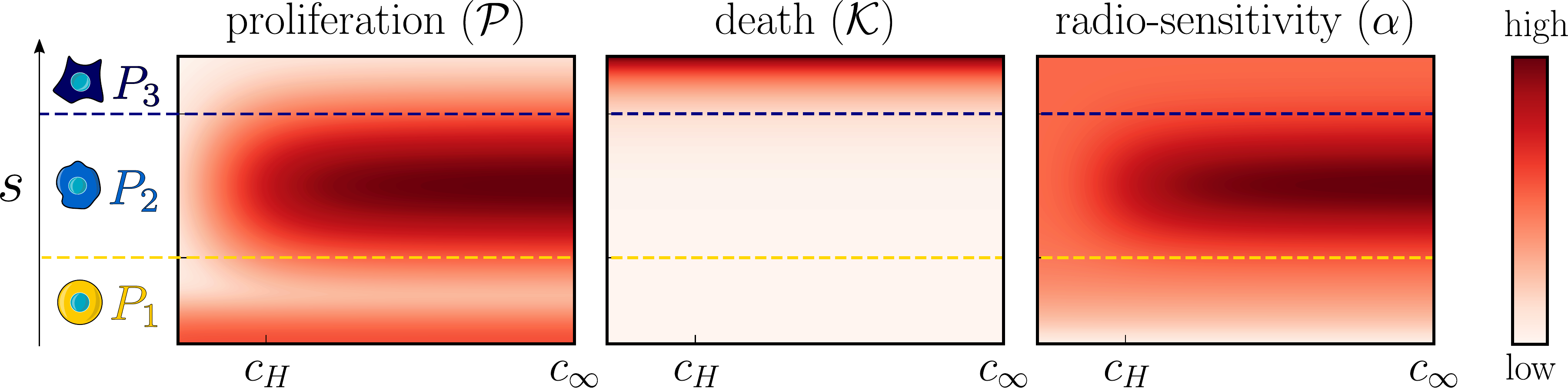}
	\caption{Illustration of the relationship between the definition of the three discrete sub-population, (namely $P_1$ (CSCs), $P_2$ (DCs) and $P_3$ (TDCs)) and the cells maximum proliferation potential (\emph{i.e.}, $\mathcal{P}$), natural death (\emph{i.e.}, $\mathcal{K}$) radio-sensitivity (\emph{i.e.}, $\alpha$). The yellow and blue dotted line indicate, respectively $s=s_1=0.3$ and $s=s_2=0.8$.}
	\label{fig:model_scheme2}
\end{figure}
When studying the tumour phenotypic composition, it will be convenient to group cells into three phenotypic subpopulations: $P_1$, $P_2$ and $P_3$. As shown in Fig.~\ref{fig:model_scheme2}, the subdivision is based on the cells' proliferative capacity ($\mathcal{P}$) and their sensitivity to natural cell death ($\mathcal{D}$) and treatment ($\alpha$ and $\beta$).
Cells in $P_1$ ($0\leq s\leq s_1$) are highly-resistant stem-like cells, while cells in $P_2$ ($s_1<s\leq s_2$) are highly proliferative and radio-sensitive. Finally, $P_3$ ($s_2<s\leq1$) consists of terminally-differentiated cells that have exited the cell-cycle and have a high death rate $\mathcal{D}$. For brevity, henceforth, we refer to $P_1$ as CSCs, $P_2$ as DCs and $P_3$ as TDCs. We emphasise that this definition of the sub-populations is not a ``firm distinction''; the thresholds $s_1$ and $s_2$ are chosen arbitrarily to showcase the dynamics of the respective populations and simplify interpretation of the simulation results. In particular, we note that small changes in the values of $s_1$ and $s_2$ do not significantly affect our results and/or conclusions. 

Given the solution $n(X,s,t)$ of Eqs.~(\ref{eq:rescaled_sys}), the proportion of cells in each of the three classes $P_1$, $P_2$ and $P_3$ at time $t$ is given by
\begin{equation}
\begin{aligned}
\Pi_{i}(t)=\frac{1}{M(t)}\int_{s_{i-1}}^{s_i} \int_0^1n(X,s,t) dXds, \quad i=1,2,3,
\end{aligned}\label{eq:Pi}
\end{equation}
where, without loss of generality, we fix $s_0=0$, $s_1=0.3$, $s_2=0.8$ and $s_3=1$. We note that, by definition, the three fractions are not independent since they satisfy the condition: $\sum_{m=1}^3 \Pi_m(t)\equiv 1$. 

While, the fractions $\Pi_i$ provide information about the overall tumour phenotypic composition, they lack information about the spatial location of the different sub-populations. Therefore, we also define the local cell fractions:
\begin{equation}
\begin{aligned}
\pi_{i}(t,X)=\frac{1}{M(t)}\int_{s_{i-1}}^{s_i}n(X,s,t) ds, \quad i=1,2,3,\label{eq:pi}
\end{aligned}
\end{equation}
to study the composition of cells in the different niches that forms within the tumour as a result of spatial heterogeneity in the oxygen levels.
%%%%%%%%%%%%%%%%%%%%%%%%%%%%%%
\section{Dynamics in the absence of treatment}
\label{sec:numerics_no_radio}
We start our analysis by presenting results from dynamic simulations of our tumour growth model, in the absence of treatment. These results highlight one of the key-features of our model: multi-stability.

\begin{figure}[tb]
	\centering
	\includegraphics[width=0.475\textwidth]{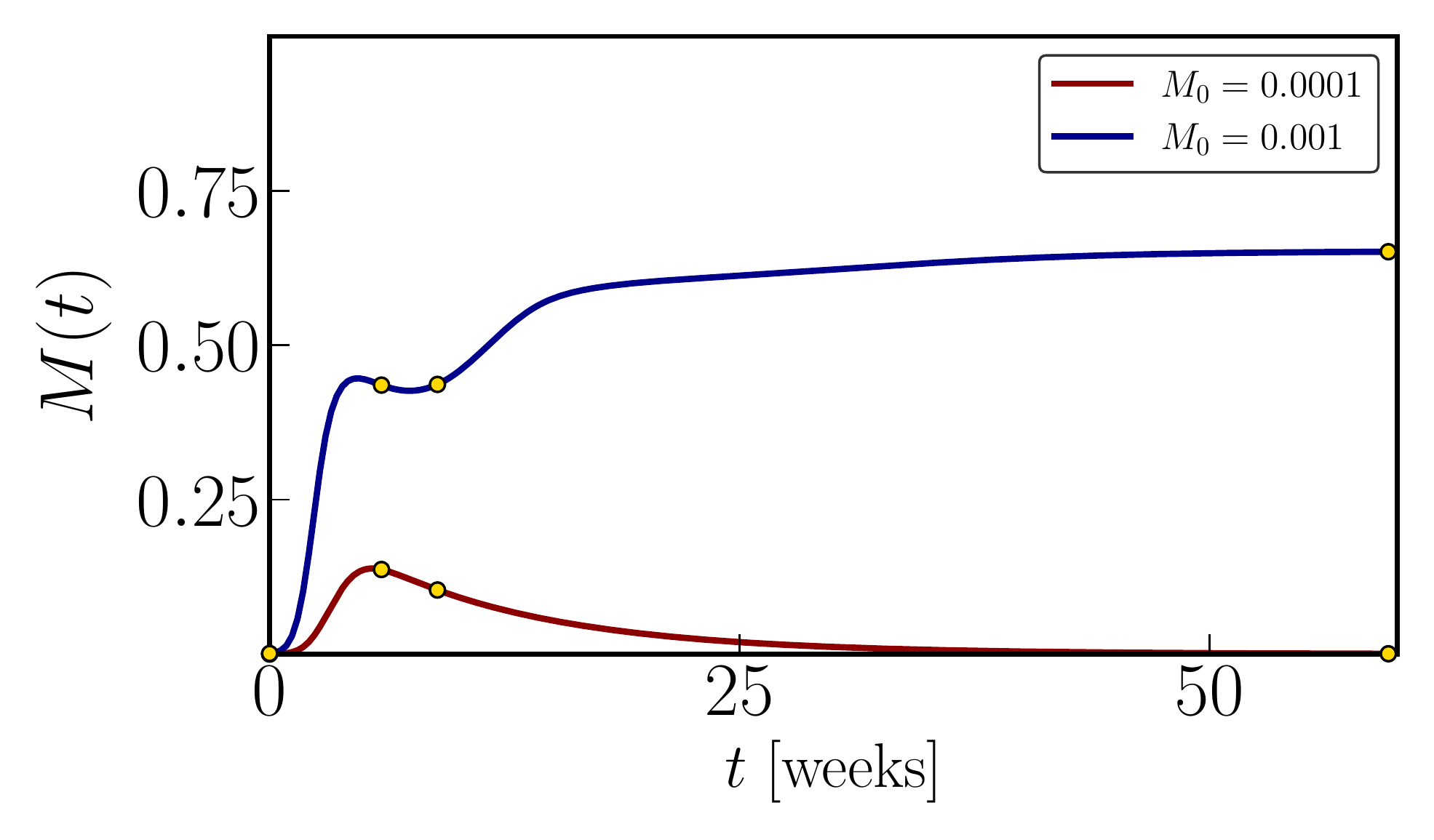}
	\vspace{-2mm}
	\caption{Numerical results generated from solving the full model, defined by Eqs.~(\ref{eq:rescaled_sys}) and (\ref{eq:initial}) with $s_0=0.5$ and two different values of $M_0$ ($M_0 = 0.001$ on the blue curve, and $M_0 = 0.0001$ on the red curve). All other model parameters are fixed at the default values specified in Tables~\ref{tab:parameters1}-\ref{tab:parameters2}, with $\gamma^*=4$ and $V_+=8\times 10^{-4}$. When $M_0 = 0.001$ (blue curve), the tumour persists and eventually evolves to a non-trivial steady state; when the initial tumour burden is decreased ($M_0 = 0.0001$, red curve), the tumour grows initially but at long times becomes extinct. 
		The yellow dots indicate the time points at which the snapshots in Fig.~\ref{fig:ex_dyn_1} are taken. }
	\label{fig:sim1totmass}
\end{figure}
\begin{figure}[ht!]
	\centering
	\begin{subfigure}{\textwidth}
		\includegraphics[width=\textwidth]{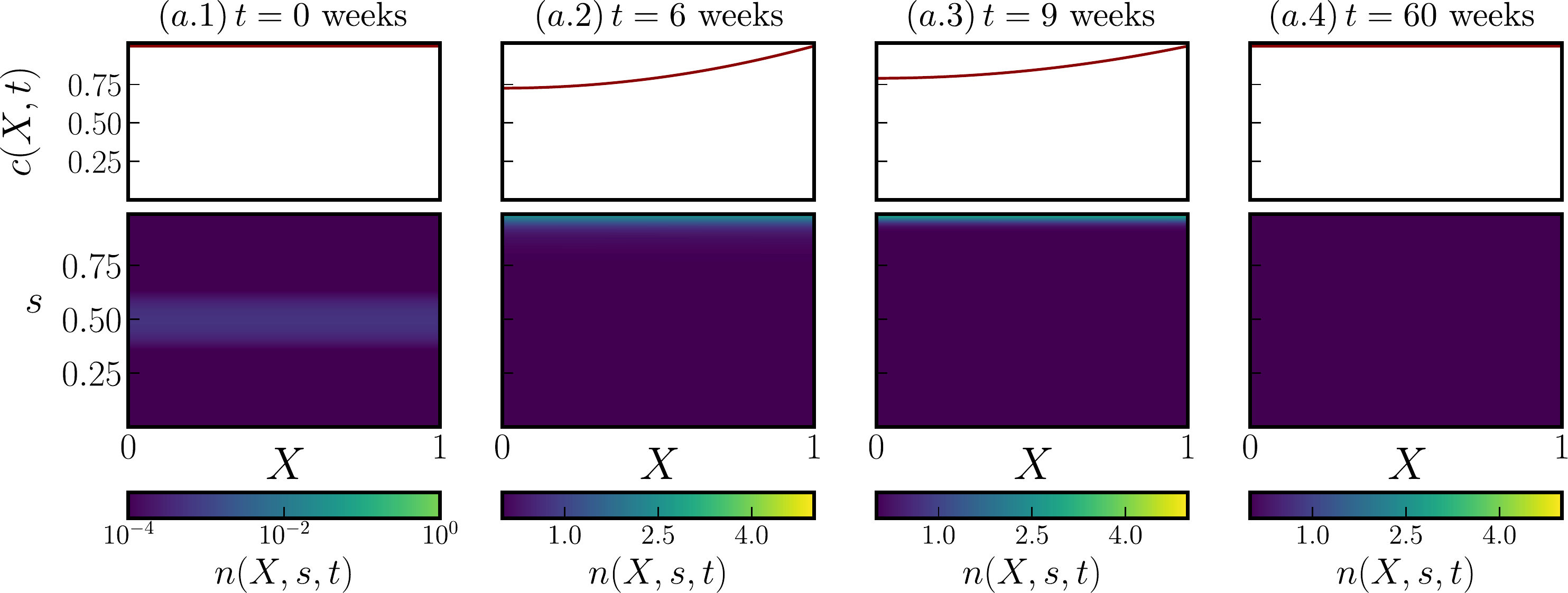}
		\caption{$M_0=0.0001$}
		\label{fig:sim1A}
	\end{subfigure}
	
	\vspace{2mm}
	\begin{subfigure}{\textwidth}
		\includegraphics[width=\textwidth]{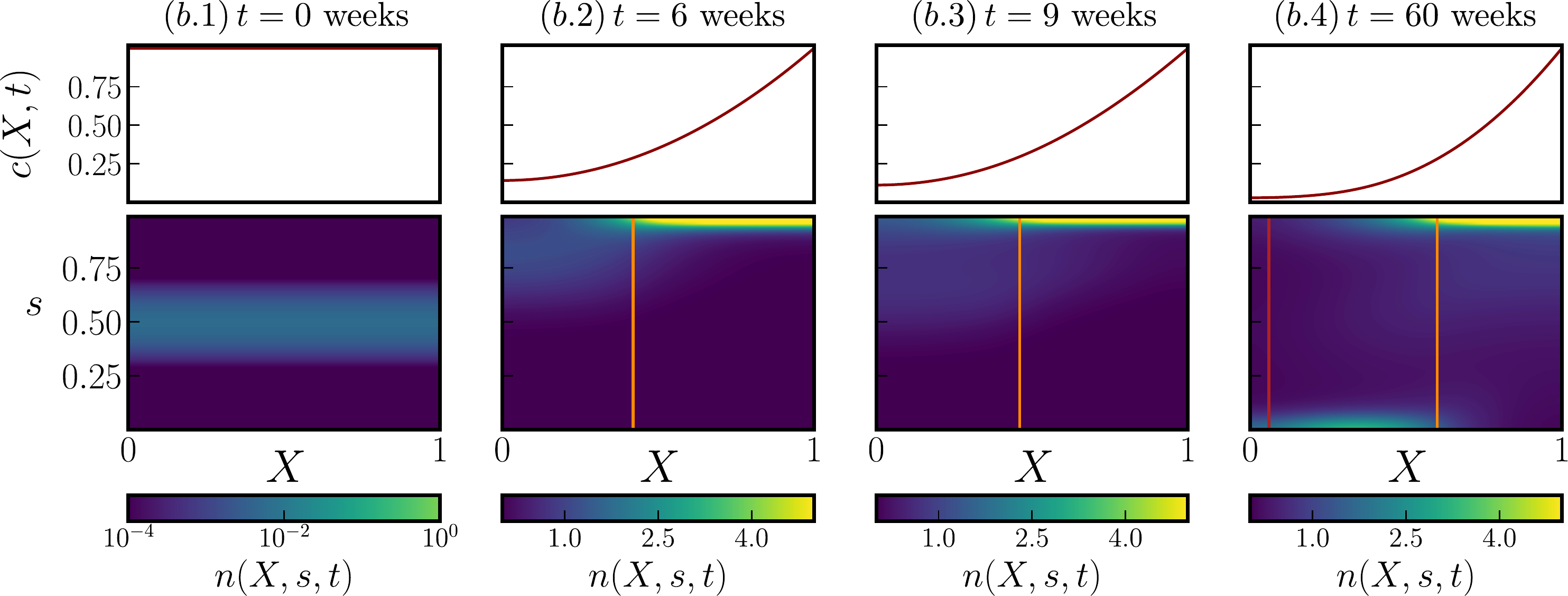}
		\caption{$M_0=0.001$}
		\label{fig:sim1B}
	\end{subfigure}	
	\caption{Time series plots showing the local cell distributions $n(x,s,t)$ for the two scenarios in Fig.~\ref{fig:sim1totmass}. Solutions are computed by numerically solving Eqs.~(\ref{eq:rescaled_sys}) with initial conditions defined by Eq.~(\ref{eq:initial}), and $s_0=0.5$. We observe long-time extinction when $M_0=0.0001$ (Fig.~\ref{fig:sim1A}) and evolution to a persistent tumour cell population when $M_0=0.001$ (Fig.~\ref{fig:sim1B}). 
		Parameter values: as per Fig.~\ref{fig:sim1totmass}. }
	\label{fig:ex_dyn_1}
\end{figure}

In Fig.~\ref{fig:sim1totmass}, we show how the tumour burden $M(t)$ evolves for two simulations in which all parameters are the same except for the initial tumour burden $M_0$. When $M_0$ is sufficiently large (e.g., $M_0 = 0.001$), the tumour invades the tissue; for smaller values of $M_0$ (e.g., $M_0 = 0.0001$) the tumour cells become extinct (see also Fig.~\ref{fig:ex_dyn_1}).

Despite the different long-time behaviour of $M$ for these two examples their initial dynamics (until about $t=9$ weeks) are qualitatively similar. In both cases, the tumour grows until $t \approx 6$ weeks. Thereafter, the tumour with the smaller initial burden, decreases monotonically towards extinction, while the tumour with the higher burden resumes growth and eventually attains a non-trivial steady state. 

To better understand why the dynamics diverge, we consider the time evolution of the cell distribution $n(X,s,t)$ (see Fig.~\ref{fig:ex_dyn_1}). In both examples, the tumour is initially composed of highly-proliferative differentiated cells thus the increase in the tumour burden $M(t)$. However, as these cells proliferate, they also become terminally differentiated. At about 6 weeks, the tumour is primarily composed of TDCs (note the large value of $n(X,1,t=6)$ in Figs.~\ref{fig:ex_dyn_1}(a.2) and \ref{fig:ex_dyn_1}(b.2)). Since the death rate of TDCs exceeds their proliferation rate (see Fig.~\ref{fig:velprol}(a)), $M(t)$ start to decrease. Further, while TDCs do not proliferate, they still consume oxygen. Consequently, if sufficiently many TDCs are present, an hypoxic region forms at the centre of the tumour (where $x\approx 0$) as observed in Fig.~\ref{fig:ex_dyn_1}(b.2). In the hypoxic \emph{niche}, cell differentiation slows down and,
if oxygen levels are sufficiently low, cells de-differentiate. The different signals experienced by tumour cells in the normoxic and hypoxic niches select for different phenotypes and drive spatial variation in the phenotypic composition of the tumour which is amplified over time (compare the cell distributions at $t=6$ and $t=60$ weeks, for example).
By contrast, in Fig.~\ref{fig:ex_dyn_1}(a.2), the number of TDCs are low so that oxygen levels remain above the hypoxic threshold $c_H$. In this case, $n(X,s,t)$ is spatially homogeneous and the tumour comprises only TDCs which are exposed to normoxia. 
As a result, the tumour is eventually driven to extinction, and the rate at which $M(t)$ decays matches $d_f$, the rate of at which TDCs die under normoxia. 

\begin{figure}[ht]
	\includegraphics[width=\textwidth]{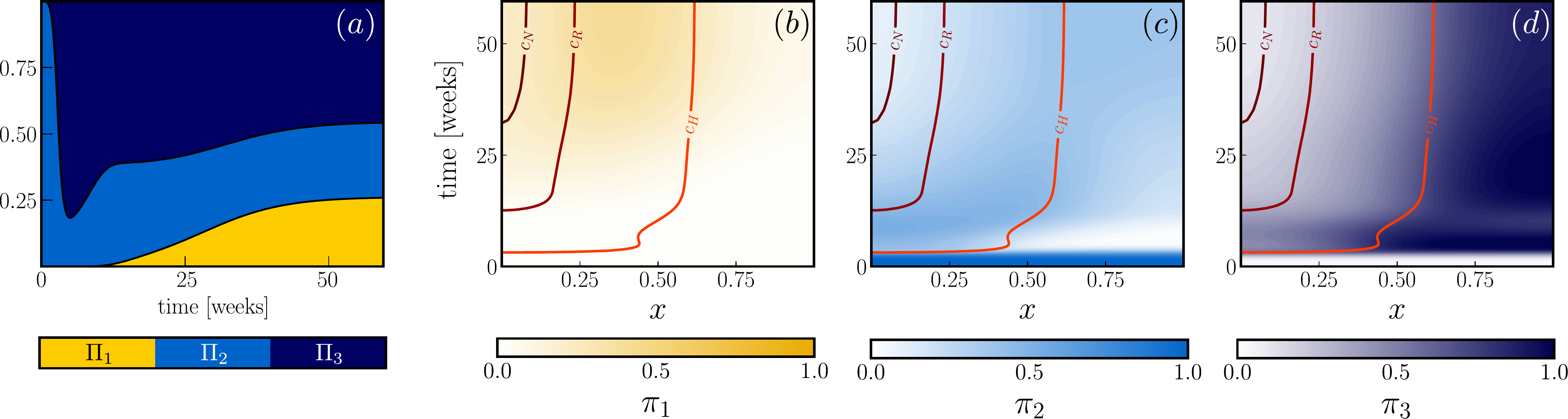}
	\caption{Series of plots showing how, for the simulation illustrated in Fig.~\ref{fig:sim1B}, the tumour composition changes over time. All parameters are fixed at the default values stated in Tables~\ref{tab:parameters1}-\ref{tab:parameters2} with $\gamma^*=4$ and $V_+=8\times 10^{-4}$ and $M_0 = 0.001$. 
		(a) Cumulative plot indicating the fractions $\Pi_i(t)$, with $i=1,2,3$, as defined in Eq.~(\ref{eq:Pi}); 
		(b)-(d) local subpopulation distribution $\pi_i(x,t)$ as defined by Eqs.~(\ref{eq:pi}). }
	\label{fig:sim1Bcomp}
\end{figure} 

Having understood the origin of the divergent behaviour observed in the two examples, we focus now on the invasion scenario (\emph{i.e}, $M_0=0.001$). As mentioned above, it is the presence of de-differentiated cells in the hypoxic region that drives secondary regrowth of $M(t)$ for $t>9$ weeks. This is apparent, when looking at Fig.~\ref{fig:sim1Bcomp}, where we show how the tumour's phenotypic composition evolves over time. When the hypoxic region first forms, the DCs localise in the hypoxic niche (see Fig.~\ref{fig:sim1Bcomp}(c)) while the normoxic region contains only TDCs (see Fig.~\ref{fig:sim1Bcomp}(d)). While DCs can proliferate, under hypoxia they stop proliferating and become quiescent. Consequently, between 6 and 9 weeks, TDCs in the outer normoxic region tend to die while DCs in the hypoxic niche survive but do not proliferate. Hence the net growth rate is negative. At later times, the DCs diffuse from the hypoxic to the normoxic region, and in the presence of increased oxygen levels they resume proliferation and drive the secondary increase in $M(t)$. This initiates a feedback loop: as $M(t)$ increases, the size of the hypoxic region also increases, creating a niche which is favourable for the development of CSCs, which drives further tumour growth. As shown in Fig.~\ref{fig:sim1Bcomp}(a), CSCs appear in the tissue only at later times ($t\approx 15$ weeks) and are localised in hypoxic regions (see $\pi_1(X,t)$ in Fig.~\ref{fig:sim1Bcomp}(b)). At long times, the fraction of CSCs, $\Pi_1(t)$, increases and asymptotes to a stationary value of $\approx 25\%$. The TDCs remain the dominant cell phenotype, comprising $\approx 50\%$ of the tumour cells. While CSCs and TDCs are located in distinct tumour regions (the hypoxic and normoxic niches, respectively), DCs eventually spread uniformly across the tissue (Fig.~\ref{fig:sim1Bcomp}(c)). Therefore, our model predicts coexistence of different cell populations not only within the same tumour, due to the formation of different niches, but also at the same spatial location, due to the cells' ability to move. The interplay between these two effects sustains tumour heterogeneity and is what eventually drives not only the formation but also the growth of the tumour. 
%%%%%%%%%%%%%%%%%%%%%%%%%%%%%%%

\section{Equilibrium states in the absence of treatment and their stability.}
\label{sec:bifurcation}
The numerical results presented in Fig.~\ref{fig:sim1totmass} suggest that the model admits multiple stable steady state solutions, at least for some parameter sets. Consequently, the long time behaviour of the system depends on the initial conditions, and their positioning with respect to the
basins of attraction of the stable steady states. 
To investigate the possible model outcomes,
we compute the steady states $(n_{\infty}(X,s),c_{\infty}(X))$ of the full non-dimensional model, Eqs.~(\ref{eq:rescaled_sys})-(\ref{eq:rescaled_sysOx1}), and study their stability. We note that, in so doing, we consider the full, time-dependent model, with no quasi-steady assumption for the oxygen.

\begin{figure}[ht]
	\centering
	\includegraphics[width=0.475\textwidth]{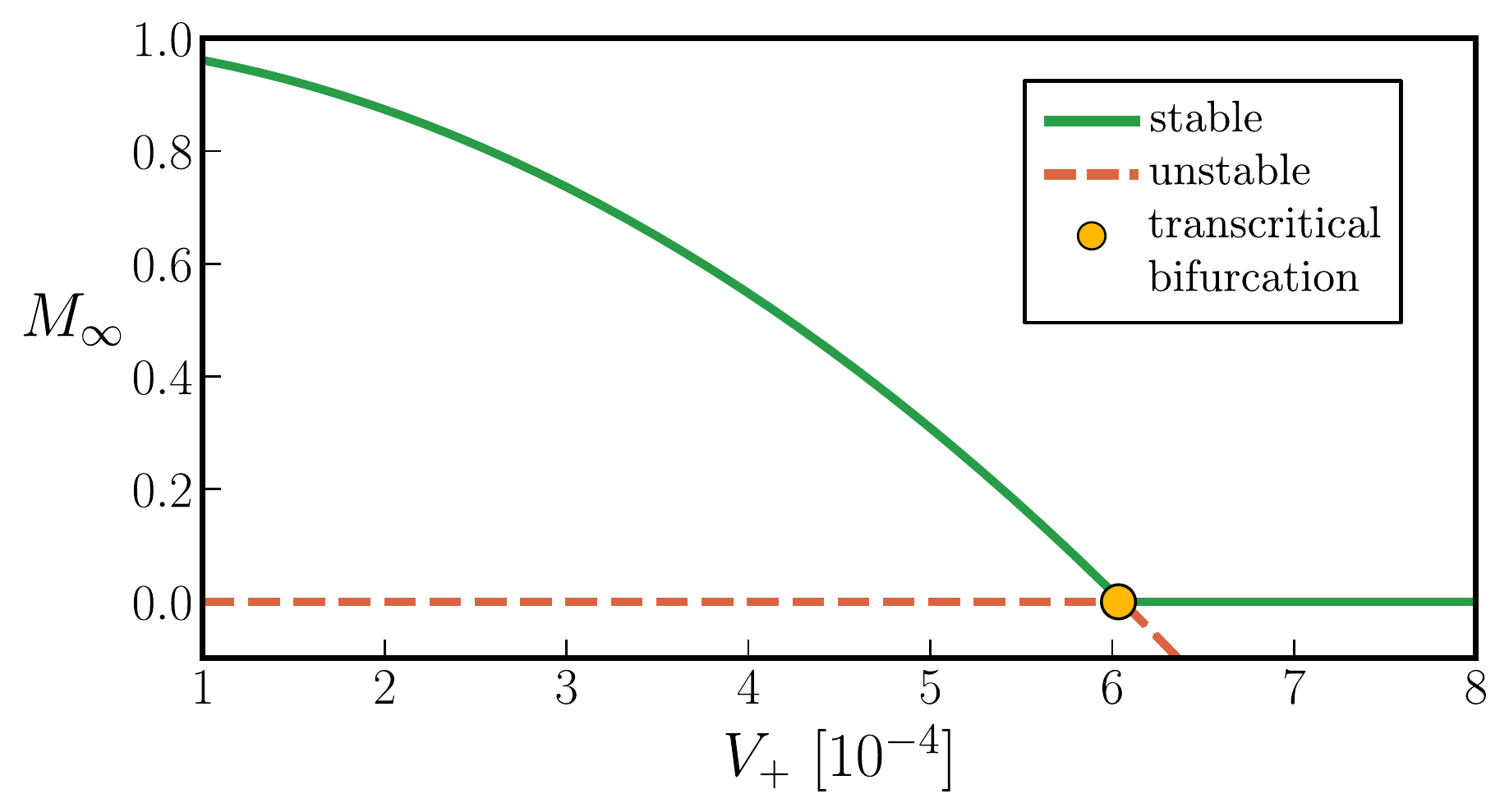}
	\caption{Bifurcation diagram for the system~(\ref{eq:rescaled_sys}) as a function of the parameter $\gamma^*$ for fixed value of $V_+$. We differentiate between stable (solid) and unstable (dashed)
		branches of the diagram and highlight with a circle the
		relevant bifurcation point.}
	\label{fig:bifwellmix}
\end{figure}

It is straightforward to show that Eqs.~(\ref{eq:rescaled_sys})-(\ref{eq:rescaled_sysOx1}) admit a unique homogeneous steady state (independent of $s$ and $X$). This is the trivial solution, with $n_\infty\equiv0$ and $c_\infty\equiv1$, and corresponds to the ``tumour extinction'' (disease-free) scenario. In~\ref{App:LSA}, we show that its linear stability can be determined from that of the simpler, well-mixed model formulation of the model which was investigated in~\cite{CELORA2021}. This is because, when decomposing the spatial perturbation into Fourier modes with wavenumber $\omega$, the particular wavenumber $\omega=0$ (corresponding, effectively, to the well-mixed case) is present. Further, since it is the most unstable wavenumber, it determines the stability of the trivial steady state. The well-mixed model is obtained by neglecting spatial variation in Eqs.~(\ref{eq:rescaled_sys}), i.e., setting $\parD{xn}$, and considering an homogeneous oxygen distribution $c\equiv c_\infty$. While all other model parameters influence the stability of the trivial steady state, we focus on parameters associated with the advection velocity $v_s$ and fix all other parameters at values based on estimates from the literature (see Tables~\ref{tab:parameters1}-\ref{tab:parameters2}). Since $c\equiv 1$ when spatial heterogeneity is neglected, we have that $v_s(s)\approx v_s^+(s)$. As shown in Fig.~\ref{fig:bifwellmix}, when $0 < V_+ \ll 1$, the trivial steady state is unstable and a non-trivial stable solution exists (see the upper green branch). As $V_+$ increases , $M_\infty$ decreases until $V_+$ crosses the critical point $V_+^{(cr)}\approx 6.06\times10^{-4}$ where the two solution branches cross and exchange stability via a transcritical bifurcation. For $V>V_+^{(cr)}$ the trivial steady state is the unique, non-negative (i.e., physically-realistic) solution and it is stable. 

The bifurcation diagram becomes more complex when we account for spatial heterogeneity. However, this will not affect the stability of the trivial steady state. In other words, the location of the critical point $V_+^{(cr)}$ is independent of $\parD{xn}$ or the parameter associated to the oxygen distribution, i.e., $\gamma_*$ and $\parD{xc}$.

\subsection{Bifurcation diagrams: bistability}
Computing heterogeneous stationary steady state solutions is challenging and, therefore, we rely on numerical continuation techniques. In particular, we use \texttt{Julia}'s package \texttt{BifurcationKit}~\cite{veltz2020} to compute solutions, study their stability and identify bifurcation points (BifurcationKit is an open-source software package for computing bifurcation diagrams). More details about the numerical techniques used can be found in~\ref{App:numerics}. 
Before continuing, we note that we present below
all of the positive steady state that were identified. These solutions were also observed when we performed dynamic numerical simulations; however we cannot, a priori, exclude the possibility 
that other stationary steady state solutions, not detected by our analysis, or other type of attractors (such as limit cycles) might exist.

As mentioned above, the parameter $V_+$ plays a critical role in the long time system dynamics. We now show how the diagram in Fig.~\ref{fig:bifwellmix} changes when we account for spatial structure. In doing so, we compute bifurcation diagrams for different values of the dimensionless oxygen consumption rate $\gamma^*$, which affects the equilibrium oxygen distribution. 

\begin{figure}[tb]
	\includegraphics[width=\textwidth]{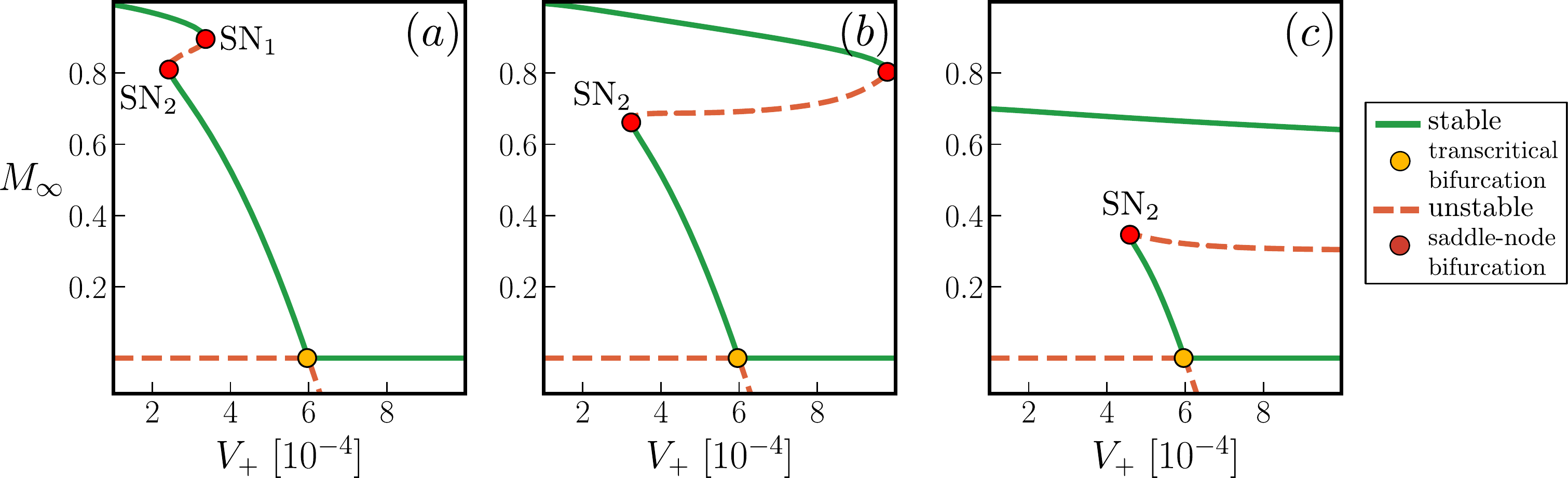}
	\caption{Series of diagrams showing how the bifurcation structure of system~(\ref{eq:rescaled_sys}) changes as the parameter $V_+$ varies, for three fixed values of $\gamma^*$: (a) $\gamma^*$ = 1.65, (b) $\gamma^*$ =2.00 and (c) $\gamma^*$ =4.00. We distinguish between stable (solid) and unstable (dashed) solution branches and highlight bifurcation points with circles.}
	\label{fig:bif_diag1}
\end{figure}

The results presented in Fig.~\ref{fig:bif_diag1} show how, as $\gamma^*$ increases through the critical value $\gamma^{(cr)}\approx 1.62$, the qualitative properties of the bifurcation diagram change. 
Comparison of Fig.~\ref{fig:bif_diag1} with the bifurcation diagram for the well-mixed scenario (i.e., Fig.~\ref{fig:bifwellmix}), we observe new steady state solutions and new bifurcation points. In particular, the two saddle-node bifurcation points, $\sn{1}$ and $\sn{2}$, define the boundary of a bistable region. Focusing on Fig.~\ref{fig:bif_diag1}(a), we see that, for small $V_+$, there exists a single stable steady state, which is characterised by a large tumour burden. As $V_+$ increases, the system undergoes a saddle node bifurcation ($\sn{2}$). For larger values of $V_+$, the system admits four non-negative steady state solutions; two of these solutions are stable and characterised by non-zero tumour burdens. As $V_+$ increases further, there is a second saddle-node bifurcation ($\sn{1}$), beyond which a unique, non-trivial, stable steady state exists whose mass decreases with $V_+$ until it crosses the trivial branch when $V_+=V_+^{(cr)}$ (the yellow circle in Fig.~\ref{fig:bif_diag1}(a)). Beyond the transcritical bifurcation, the trivial steady state (with $M_\infty =0$) is the only stable solution. For larger values of $\gamma^*$ (see Figs.~\ref{fig:bif_diag1}(b) and~\ref{fig:bif_diag1}(c)), the tumour burden on the upper-most branch decreases 
and $\sn{1}$ moves to the right, passing beyond $V_+^{(cr)}$. Note that in Fig.~\ref{fig:bif_diag1}(c), $\sn{1}$ does not appear as it is located beyond the interval under
consideration, $V_+\in[10^{-4},10^{-3}]$. The second saddle-node bifurcation point, $\sn{2}$, also moves to the right (i.e. higher $V_+$) but it is less sensitive to changes in $\gamma^*$. In particular, comparing Fig.~\ref{fig:bif_diag1}(b) and Fig.~\ref{fig:bif_diag1}(c), we see that $\sn{2}$ approaches the transcritical bifurcation point. Further numerical analysis of the fold points (see Fig.~\ref{fig:foldcont} in~\ref{app:foldcont}) suggests that the transcritical and fold bifurcations intersect at infinity (i.e., $\sn{2}(\gamma^*)\rightarrow V_+^{(cr)}$ as $\gamma^*\rightarrow\infty$).

\begin{figure}[htb]
	\includegraphics[width=\textwidth]{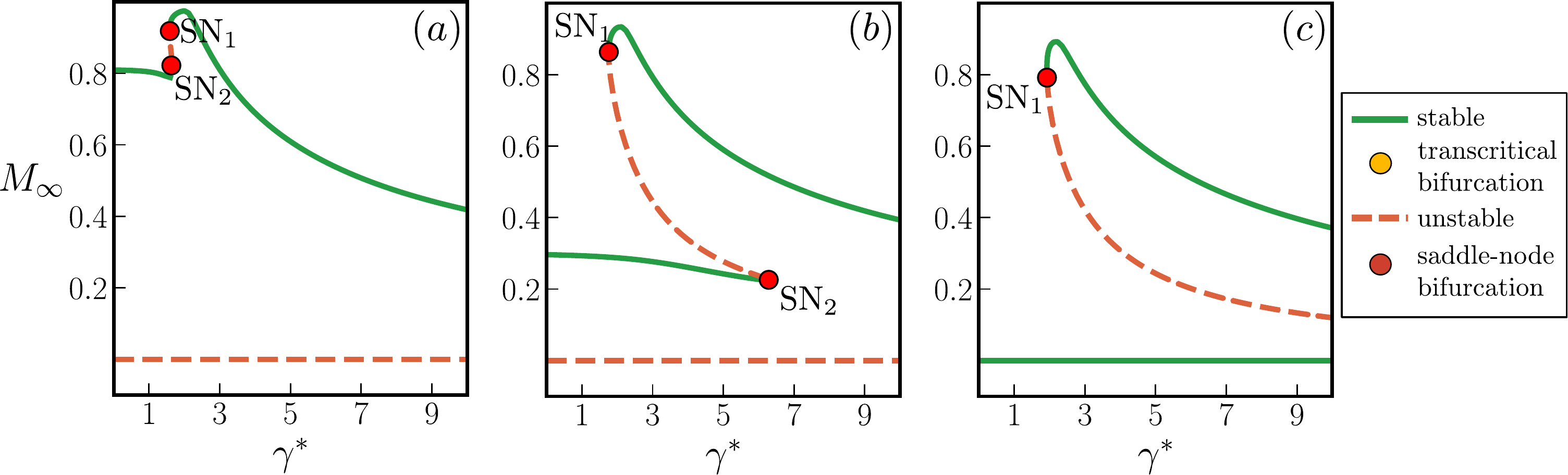}
	\caption{Series of bifurcation diagrams of the system~(\ref{eq:rescaled_sys}) with $\gamma^*$ being our bifurcation parameter. We compute the diagrams for three fixed values of $V_+$: (a) $V_+=2.5\times 10^{-4}$, (b) $V_+=5\times 10^{-4}$ and (c) $V_+=8\times10^{-4}$. We distinguish between stable (solid) and unstable (dashed) branches of solutions and highlight with circles the bifurcation points.}
	\label{fig:bif_diag2}
\end{figure}

To better understand the role played by $\gamma^*$, we fix $V_+$ and compute the system steady states, viewing $\gamma^*$ as our continuation parameter. The corresponding  diagrams, presented in Fig.~\ref{fig:bif_diag2}, show that for slow advection velocities (i.e., $V_+=2.5\times10^{-4}$), the bi-stability region is narrow, and the unstable branch connecting the two fold points is almost vertical. As $V_+$ increases to $5\times 10^{-4}$ (see Fig.~\ref{fig:bif_diag2}(b)), $\sn{2}$ shifts towards the bottom-right of the diagram and the size of the bistable region increases. The position of $\sn{1}$ moves in a similar manner, but the effect is less pronounced. We note that the upper stable branch is non-monotonic, with $M_\infty$ initially increasing past the supercritical saddle-node point ($\sn{1}$) and subsequently decreasing as $\gamma^*$ increases further. This contrasts with the lower stable branch, which decreases monotonically until it disappears at the saddle-node bifurcation point
$\sn{2}$. In Fig.~\ref{fig:bif_diag2}(c), we take
$V_+>V_+^{(cr)}$ (i.e., $V_+$ is beyond the transcritical bifurcation point shown in  Fig.~\ref{fig:bif_diag1}). Comparing Figs.~\ref{fig:bif_diag2}(b) and.~\ref{fig:bif_diag2}(c), we see that
the middle stable branch disappears and the trivial steady state becomes stable. This raises the question of what happens to the second saddle-node point ($\sn{2}$). As $V_+\rightarrow V_+^{(cr)}$ %and $V_+< V_+^{(cr)}$, 
from below, $\sn{2}$ shifts to the right ($\sn{2}(\gamma^*)\rightarrow \infty$). Since $\sn{2}$ becomes infinite at a finite value of $V_+$, our analysis suggests that the transcritical bifurcation and $\sn{2}$  annihilate each other at infinity, i.e., at $\gamma^* \rightarrow +\infty$. This is validated by using fold continuation methods which follow the locations of $\sn{1}$ and $\sn{2}$ in the ($V_+$,$\gamma^*$) plane (the results are presented in~\ref{app:foldcont}). Consequently, in Fig.~\ref{fig:bif_diag2}(c), where $V_+>V_+^{(cr)}$, the bifurcation diagram simplifies. For $0 < \gamma^* < 1.93$, extinction is the only possible long term solution; however as $\gamma^*$ increases beyond $\sn{1}$ (i.e., $\gamma^*> 1.93$), two outcomes are possible: either the tumour goes extinct, or it persists, giving rise to what we term the ``tumour-invasion'' steady state. 

\begin{figure}[htp]
	\centering
	\includegraphics[width=0.95\textwidth]{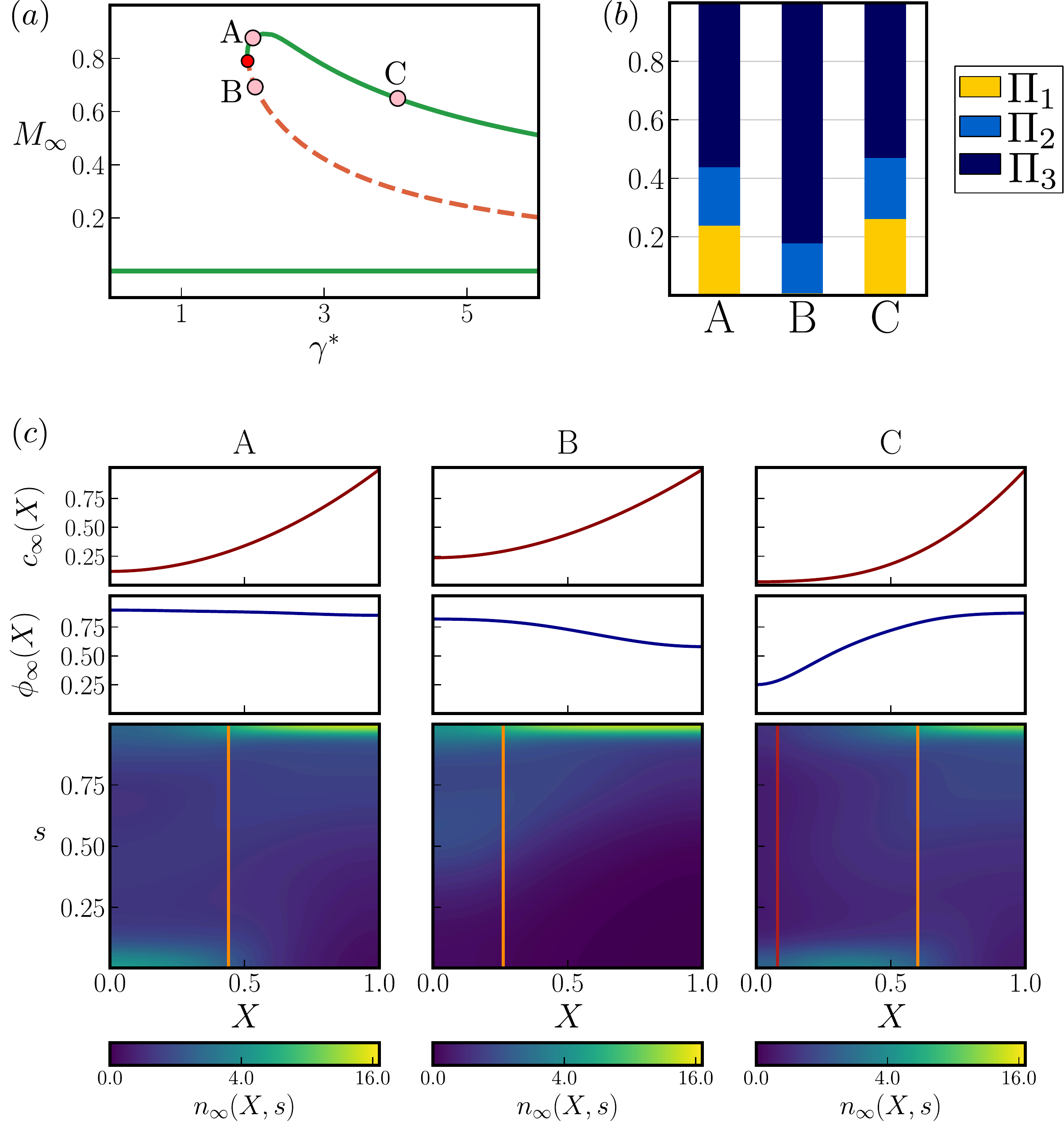}
	\caption{Characterization of the steady state solution of system~(\ref{eq:rescaled_sys}). (a) We reproduce the bifurcation diagram from Fig.~ \ref{fig:bif_diag2}(c) and indicate the equilibrium points of interest (A,B,C); (b) we compare the tumour phenotypic composition ($\Pi_i$ see Eq.~(\ref{eq:Pi})) at points A, B, and C; (c) we illustrate the equilibrium solution, i.e., oxygen profile $c_\infty$, re-scaled cell number density $\phi_\infty$, and the local distribution of cell phenotype $n_\infty(X,s)$. In the surface plot for $n$, the vertical orange and red lines indicate 
		the boundaries of the  hypoxic ($X_H$) and necrotic regions ($X_N$), respectively. The parameter used to generate these results are as follows: $V_+=8\times 10^{-4}$ and $\gamma^*=2.0$ (A,B) or $\gamma^*=4.0$ (C).}
	\label{fig:branches_car}
\end{figure} 

\paragraph{Characterisation of the computed equilibria}

Thus far, we have identified the number of steady state solutions of Eqs.~(\ref{eq:rescaled_sys}) and determined their stability. 
We now characterise these solutions by focusing on a couple of characteristic examples. In particular in Fig.~\ref{fig:branches_car}, we compare the cases of low ($\gamma^*=2.0$) vs high ($\gamma^*=4.0$) oxygen consumption rates, while fixing $V_+=8.0\times10^{-4}$. When $\gamma^*=2.0$ (i.e., just to the right of the critical saddle node bifurcation $\sn{1}$, see red dot in Fig.~\ref{fig:branches_car}(a)), equilibria A and B have a similar tumour burden, $M_{\infty}$, but their internal phenotypic compositions differ significantly (see Fig.~\ref{fig:branches_car}(b)). While approximately $20\%$ of cells for the stable steady state A are stem-like, CSCs are completely absent from the unstable steady state B. Comparison of the spatial distributions (see Fig.~\ref{fig:branches_car}(c)), reveals that steady state A is characterised by a large hypoxic niche in which CSCs accumulate (here $\min_X c_\infty(X) < c_H$). While steady state B also possesses an hypoxic region (here mild hypoxia as $\min_X c_\infty(X) \approx c_H$), this is smaller and does not contain CSCs; we find instead coexistence of TDCs and DCs in the hypoxic niche, where DCs are uniquely localised in the hypoxic region and absent from the normoxic tissue regions. Plots of  $\phi_{\infty}(X)$, indicate that the rescaled cell density is almost constant for steady state A, whereas for B it increases towards the tumour interior (where $X\approx 0$). This is because cell growth is concentrated in the hypoxic niche and dominated by slowly proliferating DCs. 

Focusing on the upper stable branch in Fig.~\ref{fig:branches_car}(a), we see that the tumour burden, $M_\infty$, decreases monotonically for sufficiently large values of $\gamma^*$. This is due to the formation of a necrotic region in which cells die (see solution C). Overall, the tumour phenotypic composition for steady state C is similar to that of A (see Fig.~\ref{fig:branches_car}(b)), with steady state C containing a slightly higher fraction of CSCs. Nonetheless, there is a marked difference in the equilibrium cell densities $\phi_{\infty}(X)$. For steady state C the maximum value of $\phi_\infty$ occurs at the oxygenated boundary ($X \approx 1$) and it decreases monotonically towards the interior, where the oxygen levels drops below $c_N$ (i.e., the necrotic region). This suggests that, in our model, while necrosis decreases the overall tumour burden, it does not significantly impact the phenotypic heterogeneity of the tumour. However, as we will show in the next sections, the presence of a necrotic region in the tumour still correlates with poor treatment outcomes in contrast to a scenario like A, where only hypoxic niches are present. 

Our analysis pertains to the local stability of the different steady states of the system. From a mathematical point of view, it is possible to drive the tumour to extinction, if the trivial solution is stable (as for $V_+=8\times10^{-4}$) and if the system enters the basin of attraction of the trivial steady state. The question of interest is, therefore, whether it is possible to apply treatment, specifically radiotherapy, in a way that modifies the long time system dynamics and drives the tumour to extinction. Given the properties of the unstable solution branch (see B in Fig.~\ref{fig:branches_car}), we might expect that the transition between the ``tumour-invasion'' and ``tumour-extinction'' regimes is mediated by the formation of a sufficiently large hypoxic region. Even if resistant CSCs are present, they would differentiate under normoxia. We conclude that a successful treatment should avoid the formation of CSC-favourable niches (i.e, hypoxia). In other words, cell numbers should be maintained at low enough levels to prevent the formation of steep oxygen gradients. 

\section{Dynamic simulations: switching between basins of attraction}
\label{sec:dyn_sim}
Before introducing treatment with radiotherapy into the model, we pause to investigate how the initial composition of the tumour affects its long time behaviour. As in Fig.~\ref{fig:branches_car}, we fix $V_+=8\times 10^{-4}$ and consider two values of the oxygen consumption: $\gamma^*=2.0$ and $\gamma^*=4.0$. 
Recall that we impose initial conditions in which the cell distribution in uniform in space and normally distributed along the stemness axis (see Eq.~(\ref{eq:initial})). In what follows, we fix the variance $\sigma_s^2$ of the Gaussian so that $\sigma_s^2=0.02$ and investigate how the long time behaviour changes as the initial tumour burden $M_0$ and the mean phenotype $s_0$ vary. We restrict attention to $s_0\in[0.25,0.75]$ so that boundary effects do not distort the Gaussian profile.

\begin{figure}[htb]
	\begin{subfigure}{\textwidth}
		\centering
		\includegraphics[width=0.975\textwidth]{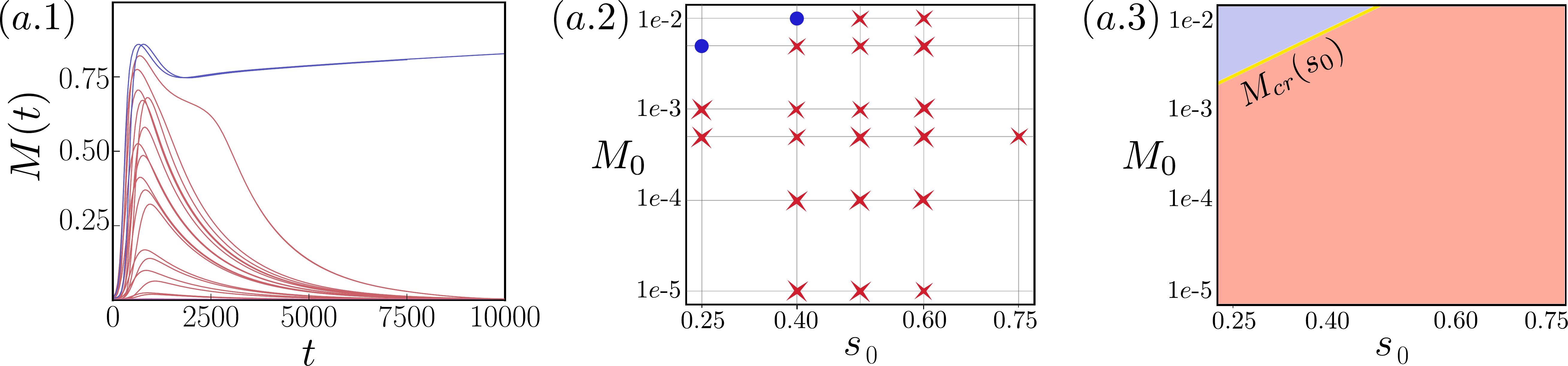}
		\caption{$\gamma^*=2.0$}
		\label{fig:evol_numerics_bifa}
	\end{subfigure}
	
	\vspace{5mm}
	\begin{subfigure}{\textwidth}
		\centering
		\includegraphics[width=0.975\textwidth]{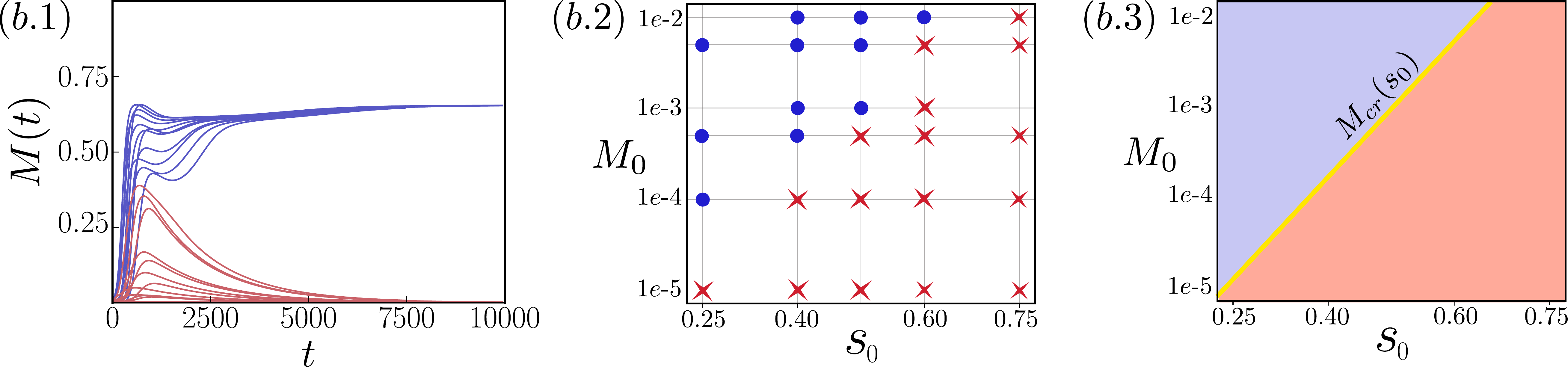}
		\caption{$\gamma^*=4.0$}
		\label{fig:evol_numerics_bifb}
	\end{subfigure}
	\caption{Series of plots showing how the long time behaviour of the tumour depends on its initial composition for the cases of (a) low and (b) high $\gamma^*$. We plot  the evolution of the tumour burden as predicted by Eqs.~(\ref{eq:rescaled_sys}) for the initial condition~(\ref{eq:initial}) (see (a.1) and (b.1)). Each curve in (a.1) (or (b.1)) corresponds to a point in the $(M_0,s_0)$ diagram (a.2) (or (b.2)); we label with red crosses initial conditions that result in tumour extinction and with blue dots the ones corresponding to 
		tumour invasion. In the subplots (a.3) and (b.3), we indicate how the $(M_0,s_0)$ plane partitions into region that map to tumour extinction (red areas) and those that map to tumour invasion (blue areas).}
	\label{fig:evol_numerics_bif}
\end{figure}

In Fig.~\ref{fig:evol_numerics_bif}, we present results from dynamic simulations. Each red curve in Fig.~\ref{fig:evol_numerics_bif}(a.1) or~\ref{fig:evol_numerics_bif}(b.1) depicts tumour extinction and corresponds to a specific cross in $(s_0,M_0)$ plane in Figs.~\ref{fig:evol_numerics_bif}(a.2) or~\ref{fig:evol_numerics_bif}(b.2); the blue curves correspond to ``tumour-invasion'' and are denoted by blue circles in Figs.~\ref{fig:evol_numerics_bif}(a.2) or~\ref{fig:evol_numerics_bif}(b.2).
From Fig.~\ref{fig:evol_numerics_bifa} 
(i.e., $\gamma^*=2.0$), we note that, for most initial conditions considered, the tumour is driven to extinction. Focusing on  Fig.~\ref{fig:evol_numerics_bif}(a.1), we see that the tumour burden $M$ does not decay monotonically; rather, it grows initially, to relatively large values ($M\approx 0.75$) in some cases, and only starts to decay at longer times. Focusing on Fig.~\ref{fig:evol_numerics_bif}(b.1), we observe again non-monotonic evolution of $M$, and the decay to extinction (see red curves) is delayed, as in Fig.~\ref{fig:evol_numerics_bif}(a.1). However, the maximum tumour burden prior to the start of the decay phase for the red curves is much smaller than those observed in Fig.~\ref{fig:evol_numerics_bif}(a.1). 
When considering the $(s_0,M_0)$ diagrams (Fig.~\ref{fig:evol_numerics_bif}(a.2) and~\ref{fig:evol_numerics_bif}(b.2)), we see that ``tumour-invasion'' scenarios are concentrated in the upper left corner. This suggests that only sufficiently large tumours, comprised of more stem-like cells ($s_0<0.5$), can persist. This effect is more pronounced for larger values of $\gamma^*$. In Fig.~\ref{fig:evol_numerics_bifb}, 
tumour survival occurs for a wider range of values of $s_0$ and $M_0$ 
than when $\gamma^*=2.0$ (Fig.~\ref{fig:evol_numerics_bifa}). 

Based on the numerical simulations, we decompose the $(s_0,M_0)$ plane into two distinct regions: one that maps to tumour-invasion (blue) and one that maps to tumour-extinction (red). As shown in Fig.~\ref{fig:evol_numerics_bif}(a.3) and (b.3), the boundary between these regions can be approximated by a straight line $M_0=M_{cr}(s_0)$ (note we are using a semi-logarithmic scale). Accordingly, for any mean phenotype $s_0$, all tumours with initial mass $M_0>M_{cr}(s_0)$ will persist. Interestingly, the model predicts that $M_{cr}$ increases with $s_0$, suggesting that as the proportion of stem-like cells ($s_0 \ll 1$) initially present increases, the smaller the value of $M_0$ for which invasion occurs. On the other hand, as $s_0\rightarrow 1$, even large tumours may be eliminated. In this case, the tumour initially comprises TDCs which are unable to drive tumour growth.  This is consistent with the idea that CSCs have the unique capacity to drive the formation and maintenance of tumour colonies. 

When applying treatment, the aim is to drive the tumour into the extinction region. Therefore, we expect that it will be easier to eradicate a tumour with a larger extinction region. From this point of view, comparing panels (a.3) and (b.3), we see that the size of the extinction region increases with $\gamma^*$, suggesting that tumours with higher consumption rates are more resilient, even though they are smaller (see Fig.~\ref{fig:branches_car}).
This is because, as the rate at which cells consume oxygen increases,  
hypoxic regions are initiated at smaller tumour volumes (i.e., smaller numbers of tumour cells), creating an environment in which CSCs can emerge and drive tumour growth (as in Fig.~\ref{fig:branches_car}, case C). On the other hand, for smaller values of $\gamma^*$, the threshold tumour burden required for the formation of hypoxia is larger, so that even large tumours may become extinct.
%%%%%%%%%%%%%%%%%%%%%%%%%%%%%%%
\section{Dynamics in the presence of treatment}
\label{sec:dynamics_with_radio}

\begin{figure}[tb]
	\centering
	\includegraphics[width=0.475\textwidth]{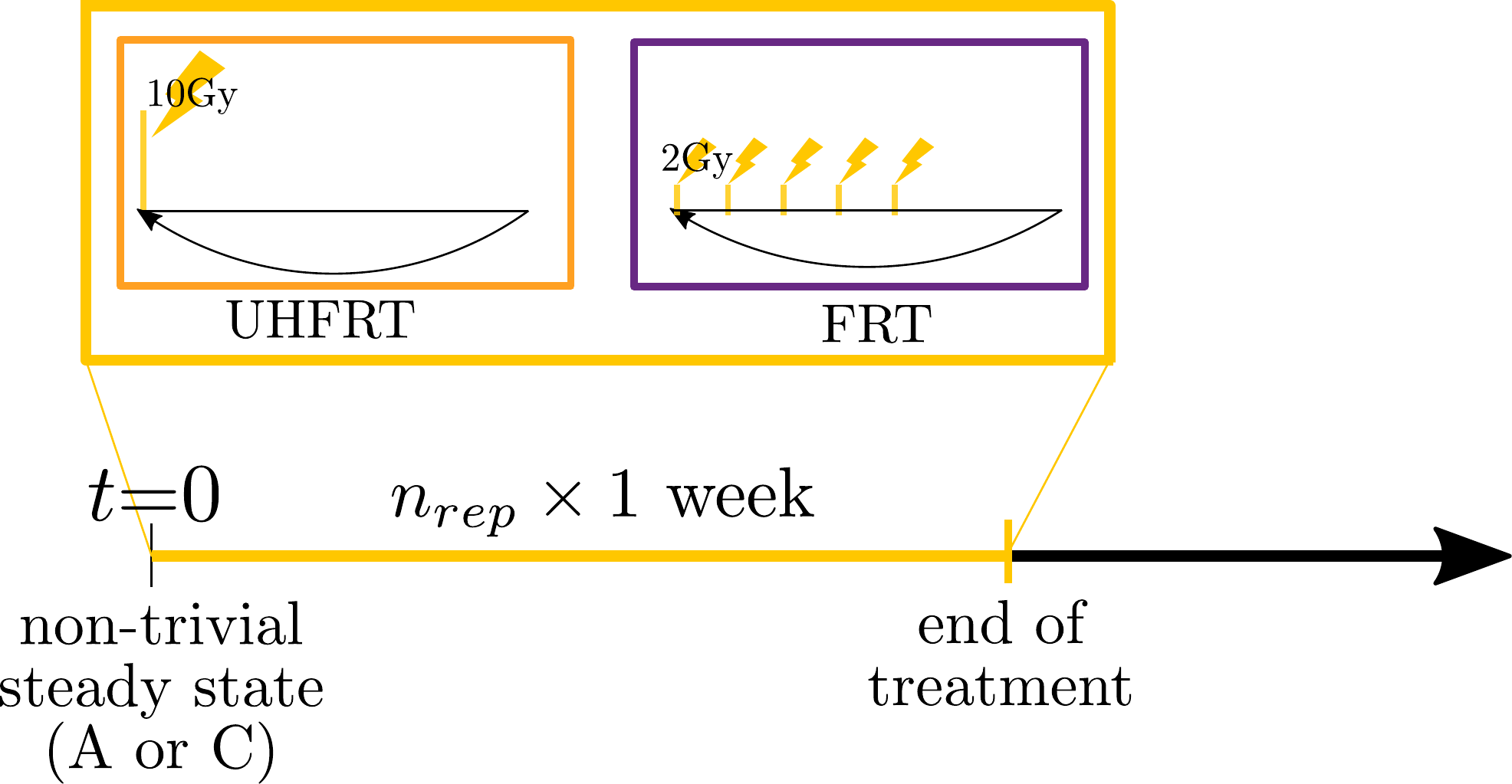}
	\caption{Schematic diagram showing how treatment is incorporated into the model. Treatment is repeated over a period of $n_{rep}$ weeks so that the final dosage delivered is $10\times n_{rep}$ Gy. We consider two different protocols: ultra-hypo-fractionated radiotherapy (UHFRT), and fractionated radiotherapy (FRT). For UHFRT, a single dose of 10 Gy is delivered on Monday of each week during treatment; for FRT, 5 doses of 2 Gy are delivered on Monday to Friday, with a break from treatment on Saturday and Sunday.}
	\label{fig:schem_treat}
\end{figure}
Having studied the model predictions in the absence of treatment, we now use it to investigate tumour responses to radiotherapy. Due to the large number of unknown model parameters, it is difficult to apply our model to patient specific data and personalised radiation scheduling. However, it can provide insight into the ways in which interactions between the different radio-resistance mechanisms determine treatment responses and, in doing so, increase understanding of the possible causes of relapse. 

As in \S\ref{sec:dyn_sim}, we fix ($V_+=8\times 10^{-4}$), so we know that tumour extinction is possible, and compare the response to treatment of tumours with different oxygen consumption rates: $\gamma^*=2.0$ and $\gamma^*=4.0$. For each value of $\gamma^*$, we consider as initial conditions the corresponding non-trivial equilibrium solutions
(i.e, steady state A or C in Fig.~\ref{fig:branches_car}). As shown in Fig.~\ref{fig:schem_treat}, we assume that the tumours receive 10 Gy of radiotherapy every week for $n_{rep}$ weeks. Treatment is delivered either as a single, high dose (Mon) or as 5 doses (Mon-Fri) with a two-day break (Sat-Sun). We refer to the first treatment strategy as ultra-hypo-fractionated radiotherapy (UHFRT) and the second one, which corresponds to standard-of-care, as fractionated radiotherapy (FRT).

\subsection{Numerical simulations}
\begin{figure}[htb!]
	\includegraphics[width=\textwidth]{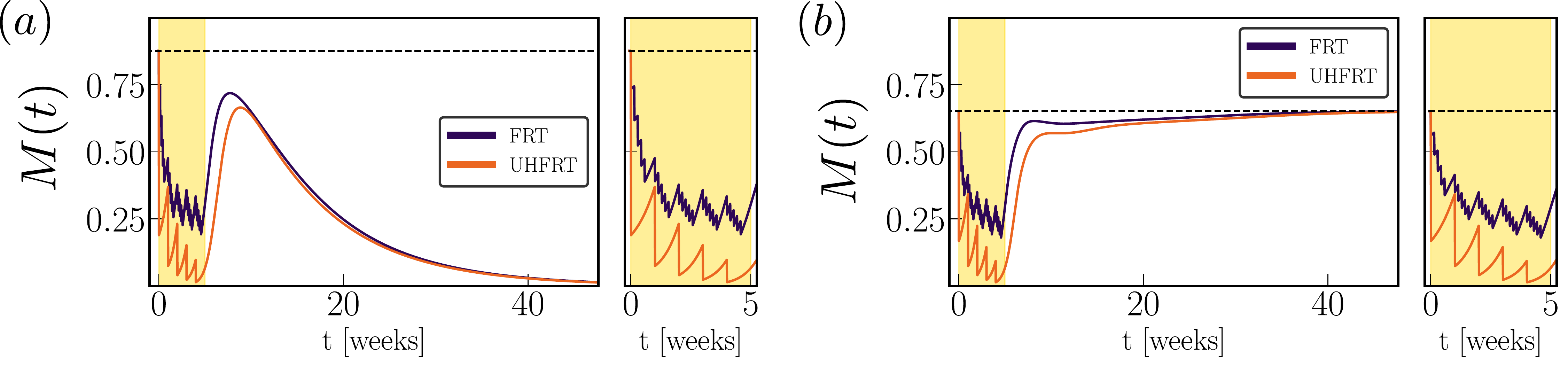}
	\caption{Evolution of the tumour burden, $M(t)$, during treatment for different values of the parameter $\gamma^*$: (a) $\gamma^*=2.0$ and (b) $\gamma^*=4.0$ . We compute $M(t)$ numerically by solving Eqs.~(\ref{eq:rescaled_sys})-(\ref{eq:c_stationary}) and applying 50 Gy radiotherapy over a period of 5 weeks according to the two protocols described in Fig.~\ref{fig:schem_treat}. The duration of treatment is highlighted by the yellow area; for each panel, we zoom on this time frame to highlight differences in the evolution of $M$. The remaining parameters are fixed at the default values listed in Tables~\ref{tab:parameters1}-\ref{tab:parameters2}. }
	\label{fig:growthcurve}
\end{figure}

In Fig.~\ref{fig:growthcurve}, we compare the evolution of the tumour burden $M(t)$ for the two treatment protocols. Fig.~\ref{fig:growthcurve}(a) shows the response of a tumour with a low rate of oxygen consumption ($\gamma^*=2.0$), while Fig.~\ref{fig:growthcurve}(b) shows the response of a tumour with a high oxygen consumption rate ($\gamma^*=4.0$). When $\gamma^*=2.0$, a 5-week course of treatment is sufficient to eradicate the tumour, regardless of the protocol used (see Fig.~\ref{fig:growthcurve}(a)). During the early stages of treatment, UHFRT reduces the tumour burden more rapidly than FRT, but after about 10 weeks, the dynamics of $M(t)$ are indistinguishable, with the tumour burden decreasing monotonically to zero in both cases. By contrast, Fig.~\ref{fig:growthcurve}(b) shows that when $\gamma^*=4.0$ treatment fails for both protocols: while the tumour initially responds to treatment, it eventually regrows to its original size. Interestingly, the tumour burdens at the end of treatment, with either FRT or UHFRT, are similar for both values of $\gamma^*$.

\begin{figure}[htb]
	\centering
	\includegraphics[width=0.95\textwidth]{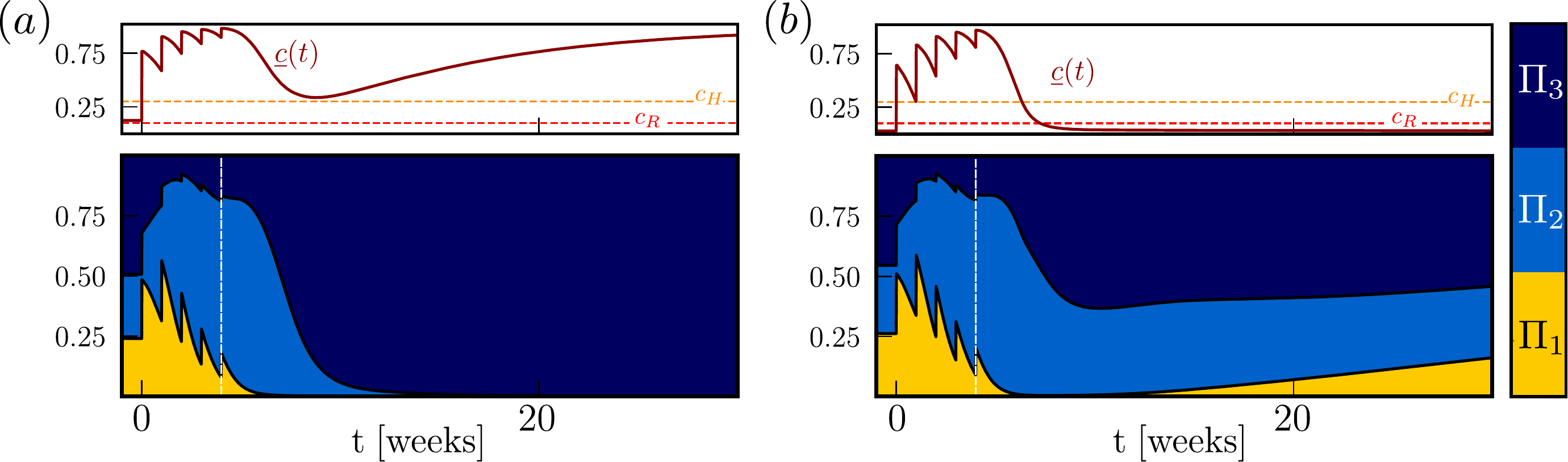}
	\caption{Numerical results showing how, for the simulations presented in Fig.~\ref{fig:growthcurve}, the tumour's internal composition changes following 5 weeks of UHFRT when (a) $\gamma^*=2.0$ and (b) $\gamma^*=4.0$. The top panels show the evolution of the minimum oxygen level within the tumour, $\underline{c}(t)=\min\limits_X c(X,t)$. The latter is compared to the oxygen thresholds $c_H$ (orange horizontal line) and $c_R$ (red horizontal line). In the lower panels, the white vertical line indicates the time at which the last dose is administered.}
	\label{fig:radio_comp_UHFRT}
\end{figure}

We now consider how the internal phenotypic composition of the tumours changes during treatment, in order to determine whether there is any early indication of different treatment outcomes. We also plot the minimum oxygen concentration recorded in the tissue $\underline{c}(t)=\min_{X}(X,t)$.
The results presented in Fig.~\ref{fig:radio_comp_UHFRT} suggest that during UHFRT the tumour composition is independent of $\gamma^*$. In both cases, RT  re-oxygenates the tumour (note that the minimum oxygen level recorded in the tissue, $\underbar{c}$, remains above $c_H$ throughout treatment). Therefore, although $\Pi_1(t)$ increases after each dose of radiotherapy (since RT selects for radio-resistant CSCs), $\Pi_1$ decreases between doses as CSCs tend to differentiate when exposed to sufficiently high oxygen levels. As a result, at $4$ weeks (i.e., just after the last dose), $\Pi_1(t=4) < \Pi_1(t=0) \approx 0.25$. By contrast, the fraction of differentiated (highly-proliferative) cells $\Pi_2(t)$ increases during treatment which explains the rapid regrowth of the tumour when treatment ends. As the number of tumour cells increases, tissue oxygen levels decrease.
While for $\gamma^*=2.0$, $\underline{c}$ remains above the threshold $c_H$, 
an hypoxic region forms at $t\approx 6$ weeks for $\gamma^*=4.0$. Only at this time point, does the internal composition of the two tumours start to diverge. While for $\gamma^*=2.0$, at long times all DCs become TDCs and the tumour is eliminated, when $\gamma^*=4.0$, DCs located in the hypoxic region de-differentiate, producing CSCs which enable the tumour to survive and continue to grow.

\begin{figure}[htb]
	\centering
	\includegraphics[width=0.925\textwidth]{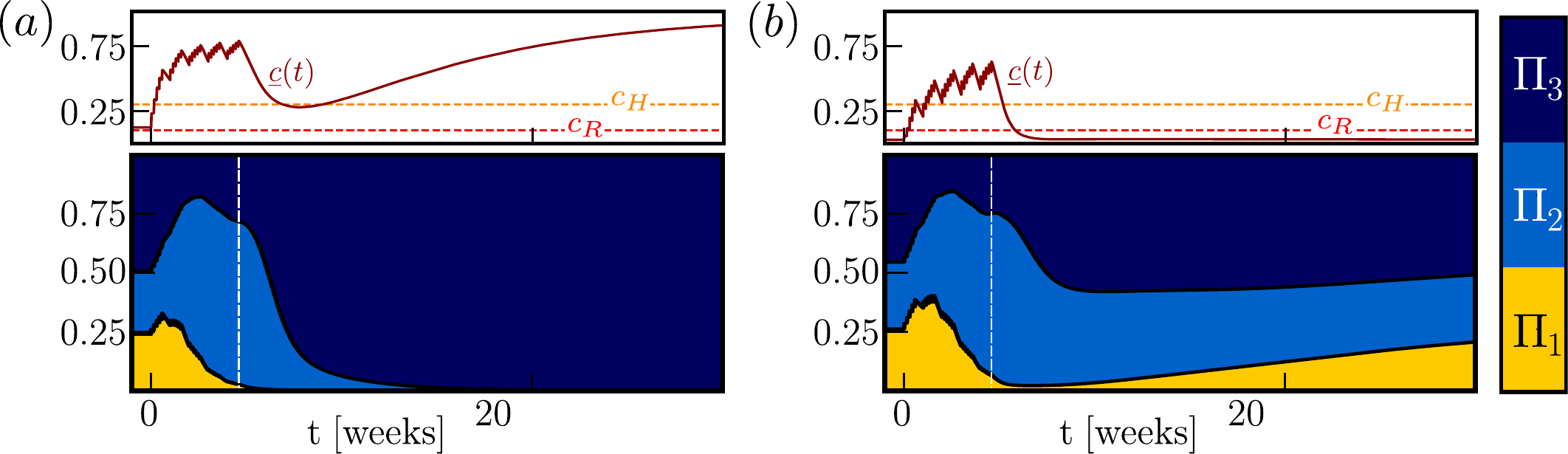}
	\caption{Numerical results showing how, for the simulations presented in Fig.~\ref{fig:growthcurve}, the tumour's internal composition changes following 5 weeks of FRT when (a) $\gamma^*=2.0$ and (b) $\gamma^*=4.0$.. In the top panels we plot the minimum oxygen level $\underline{c}$ recorded in the tissue at any time and compare it with the relevant oxygen threshold $c_H$ and $c_R$.}
	\label{fig:radio_comp_FRT}
\end{figure}

Focusing now on FRT (see Fig.~\ref{fig:radio_comp_FRT}), we observe that the evolution of $\underline{c}$ differs during the early stages of treatment for the two values of $\gamma^*$. For $\gamma^*=2.0$, one cycle of FRT is sufficient to raise oxygen levels above the hypoxic threshold $c_H$, while for $\gamma^*=4.0$ two cycles are required. As a result of differences in the oxygen levels, a larger fraction of CSCs are present in Fig.~\ref{fig:radio_comp_FRT}(b) during the initial treatment phase. However, for both scenarios, the value of $\Pi_1$ at the end of treatment (here indicated by the vertical white line) is negligible, and significantly smaller than the value recorded after the last dose for UHFRT (see Fig.~\ref{fig:radio_comp_UHFRT}). However, FRT is less effective than UHFRT in re-oxygenating the tumour because fewer cells are killed. For $\gamma^*=4.0$, this results in the rapid formation of an hypoxic region after the end of treatment so that a small fraction of CSCs persist in the tumour, in contrast to the behaviour observed in Fig.~\ref{fig:radio_comp_UHFRT}(b). 
As mentioned above, in the case of FRT, the dynamics of $\underline{c}$ changes markedly with $\gamma^*$, even during  the early phase of treatment, making tumour oxygen levels a possible biomarker for predicting outcomes in response to FRT. This is in contrast to UHFRT, where the dynamics of $\underline{c}$ (see Fig.~\ref{fig:radio_comp_UHFRT}) appears to be independent of $\gamma^*$ during the treatment period so that it can not be used as an early indicator of outcome in response to UHFRT.

The simulation results presented in this section, indicate that oxygen plays a key role in defining the response of a tumour to RT due in large part to its (eventual) impact on the dynamics of CSCs. From this viewpoint, the dimensionless oxygen consumption rate $\gamma^*$, which influences the distribution of oxygen in the tumour, plays a key role in determining the treatment outcome. Overall, we see that low values of $\gamma^*$ correlate with better responses to both UHFRT and FRT.  
Our findings support the possible benefit of combining RT and metabolic inhibitors, which can reduce oxygen consumption by tumour cells by inhibiting mitochondrial activity, as recently proposed by several experimental studies~\cite{Ashton2016,Fiorillo2016,Mudassar2020}.

\subsection{Alternative protocols: the possible benefit of waiting}
\label{sec:alternative_protocols}
The results presented in the previous section suggest that tumours with higher oxygen consumption rates are more aggressive and might not respond well to standard treatment. Here we investigate how RT scheduling might be adapted to improve treatment efficacy. Classical approaches for RT scheduling look at the total number of doses and the level of fractionation based on the concept of a {\it Biologically Effective Dose} (BED) \cite{Barendsen1982,Fowler1989}. Under the assumption that $\alpha$ and $\beta$ in the LQ model~(\ref{eq:LQ}) are constant parameters, the BED is regarded as a measure of the total biological dose delivered to the tissue for a certain dose fractionation schedule \cite{JONES2001}. However, our model illustrates how $\alpha$ and $\beta$ may change over time, as  treatment affects the tumour's internal composition and its micro-environment (i.e, oxygen levels). Consequently, the time at which a dose is administered can impact the overall outcome and must therefore be taken into consideration when planning treatment schedules. We therefore use our model to test whether waiting longer periods between treatment cycles (for example, we might repeat treatment every two weeks instead of every week) might be beneficial for more aggressive tumours ($\gamma^*=4$).

\begin{figure}[htb]
	\centering
	\includegraphics[width=0.9\textwidth]{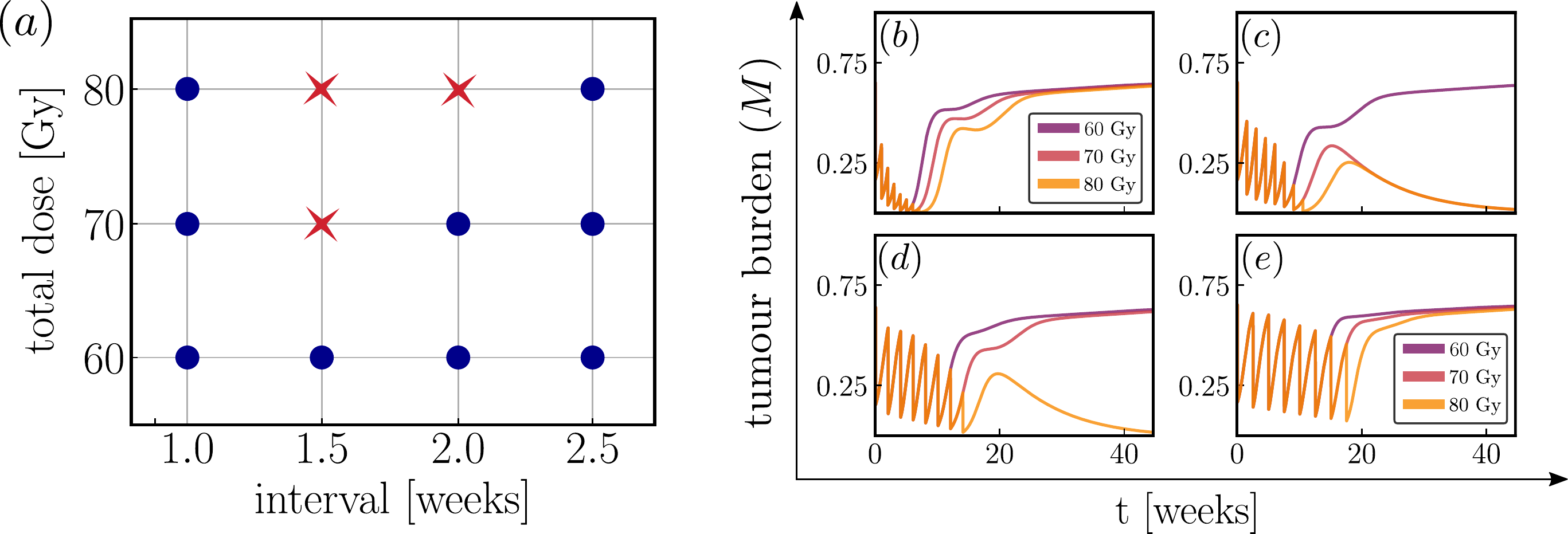}
	\caption{Comparison of different schedules for UHFRT where the interval between successive doses and the total dose given are changed. (a) Summary of the effect of different treatments (red cross: tumour goes extinct; blue circle: treatment fails and the tumour persists); 
		(b)-(e) series of plots showing how the tumour burden $M(t)$ evolves for the different scenarios in (a); while colour indicates the total dose, each subplot corresponds to a different dose interval: (b) $1$ week; (c) $1.5$ weeks; (d) $2$ weeks; (e) $2.5$ weeks.}
	\label{fig:gridUHFRT}
\end{figure} 

The results for UHFRT are summarised in Fig.~\ref{fig:gridUHFRT}. Here the interval between cycles of treatment corresponds to the interval between consecutive doses. When the interval between doses is 1 week, as in Fig.~\ref{fig:growthcurve}, treatment fails, even when we increase the total dose delivered to 80 Gy. By allowing more time between doses, we observe better responses with a total dose of 70 Gy being sufficient to prevent recurrence when there is an interval of 1.5 weeks between doses. However, if the interval between doses is too long as in Fig.~\ref{fig:gridUHFRT}(e), cells have time to grow between doses so that, even after a total dose of 80 Gy, the tumour burden $M$ is relatively large ($\approx 0.1$). Overall we observe that the shorter the interval between doses, the smaller the tumour burden at the end of treatment. Yet, recurrence is observed for inter-dose interval of 1 week and not for intervals of 1.5 weeks (the
latter is true for 70 and 80 Gy total doses). These results suggest that the different outcomes cannot be (solely) understood on the basis of the overall tumour burden.

\begin{figure}[htb]
	\centering
	\includegraphics[width=0.95\textwidth]{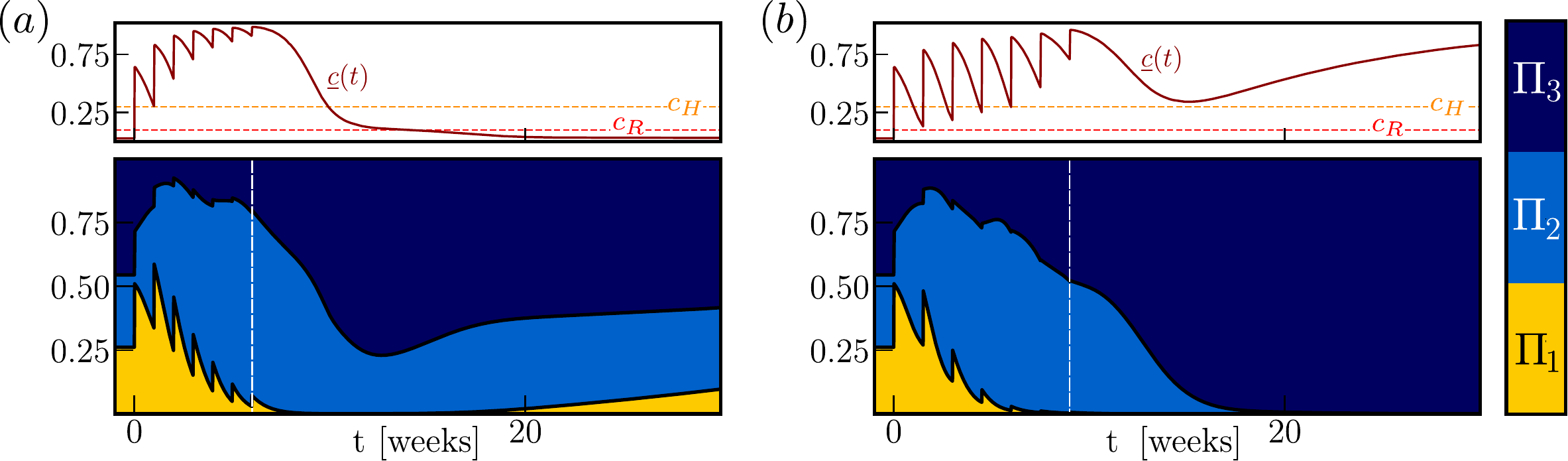}
	\caption{Plots showing how, for UHFRT, the evolution of the internal phenotypic composition of the tumour changes as the period between successive treatment cycles  is increased from (a) 1 week to (b) 1.5 weeks. We apply a total dose of 70 Gy and the vertical white line indicates the end of treatment. On the top panel we plot the evolution of the minimum oxygen level recorded in the tissue.}
	\label{fig:comp_int_UHFRT}
\end{figure} 

In Fig.~\ref{fig:comp_int_UHFRT}, we show how the internal tumour composition changes over time when a total of $70$ Gy is delivered waiting either 1 (a) or 1.5 (b) weeks between doses. While tumour re-oxygenation is less effective when waiting longer between doses, a dose interval of 1.5 weeks is more effective in reducing the overall fraction of CSCs. This is evident when comparing the values of $\Pi_1$ for the two protocols at the end of treatment (indicated by the vertical white line). Given that a single dose of 10 Gy is sufficient to reoxygenate the tissue, waiting longer  between doses gives more time for CSCs to differentiate into radio-sensitive DCs. However, if the time between doses is too long (such as 2.5 weeks), the regrowth of the tumour between doses would result in the formation of an hypoxic \emph{niche} in which CSCs would start to accumulate resulting in increased resistance of the tissue. This explains the non-monotonic dependence of treatment outcomes on inter-dose period reported in Fig.~\ref{fig:gridUHFRT}.

\begin{figure}[htb]
	\centering
	\includegraphics[width=0.9\textwidth]{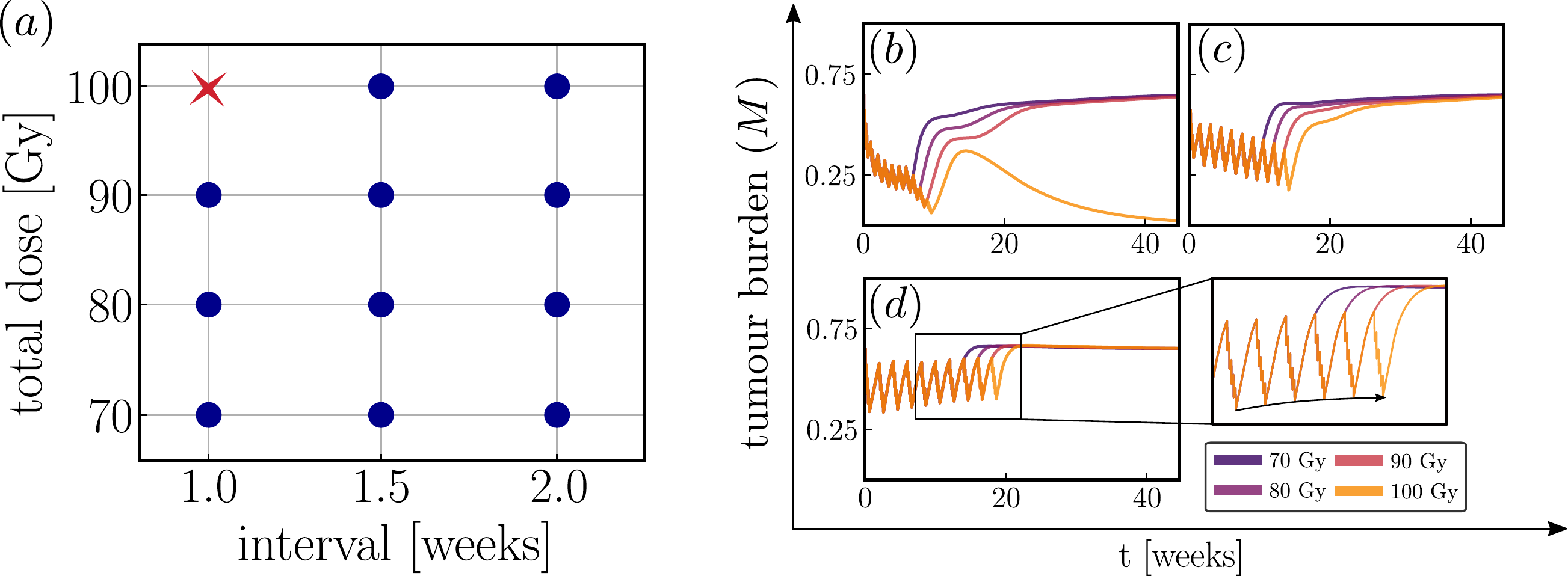}
	\caption{Comparison of different schedules for FRT where we change the interval between successive treatment cycles and the total dose given over the treatment period. (a) Summary of the effect of different treatments (red cross: tumour goes extinct; blue circle: treatment fails and tumour persists); (b)-(e) series of plots showing how the tumour burden $M(t)$ evolves for the different scenarios in (a); while colour indicates the total dose, each subplot corresponds to a different interval: (b) $1$ week; (c) $1.5$ weeks; (d) $2$ weeks. The insets in panel (d) highlight that cycles of treatment become less effective at diminishing the tumour burden due to the accumulation of radio-resistance cells (see black arrow interpolating the tumour burden at the end of each treatment cycle).}
	\label{fig:gridFRT}
\end{figure}

When we increase the interval between treatment cycles for FRT, rather than UHFRT, our model predicts a different trend. In this case, the interval between cycles corresponds to the time between the administration of the first dose of the five $2$ Gy doses, which are still delivered over 5 consecutive days. 
As shown in Fig.~\ref{fig:gridFRT}(a), the efficacy of treatment decreases monotonically with the inter-cycle interval. While for a standard interval of 1 week a total of 100 Gy is effective in preventing relapse, other schedules fail. Considering the growth curves in Fig.~\ref{fig:gridFRT}(d) (corresponding to a 2 weeks interval between cycles), we see that, by waiting longer, the tumour acquires resistance to treatment. Indeed, the value of $M$ after the last dose of each cycle increases which indicates strong selection for resistant phenotypes.

This selection is evident when we look at the evolution of the internal tumour composition n Fig.~\ref{fig:comp_int_RT}. For an interval of 1 week, oxygen levels in the tumour increase during treatment and the lack of an hypoxic niche allows CSCs to de-differentiation. As a consequence, CSCs are negligible at the end of treatment. On the contrary, for intervals of two weeks, the treatment fails to re-oxygenate the tissue, the hypoxic \emph{niche} persists and oxygen levels remain below $c_R$. As a result, large numbers of CSCs accumulate in the tissue, well above their equilibrium value ($\approx 25\%$). Consequently, subsequent rounds of RT become less effective in killing cells. Once treatment ends, the large number of CSCs in the tumour enables it to quickly evolve to a non-zero steady state.

\begin{figure}[htb!]
	\centering
	\includegraphics[width=\textwidth]{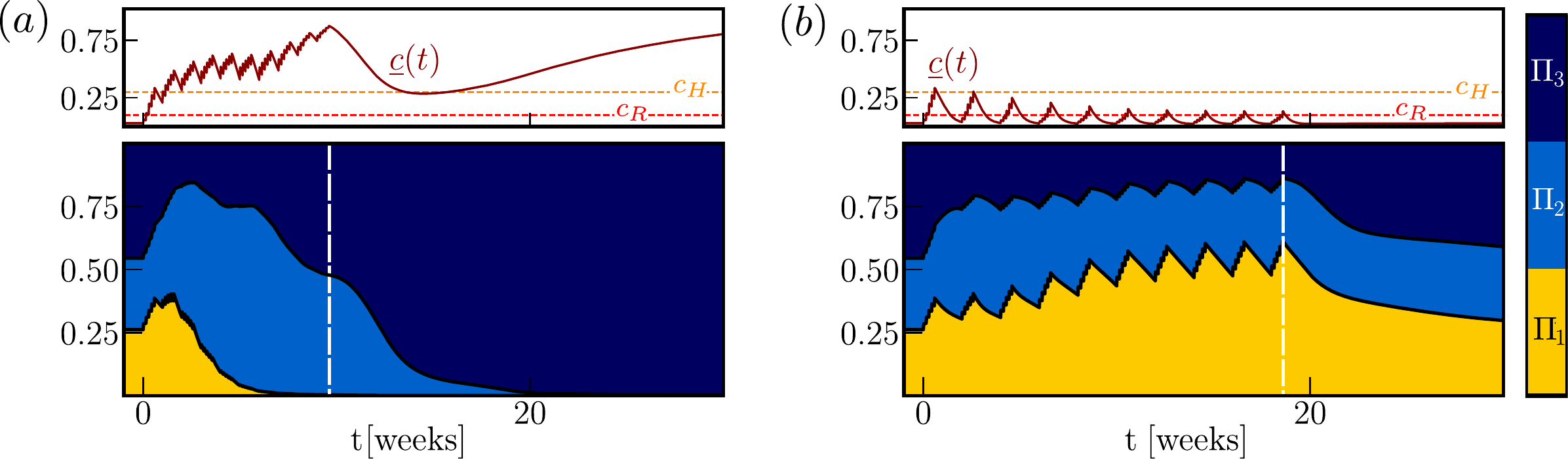}
	\caption{Plots showing how, for FRT, the evolution of the internal phenotypic composition of the tumour changes as the period between successive treatment cycles  is increased from (a) 1 week to (b) 2 weeks. We apply a total dose of 100 Gy and the vertical white line indicates the end of treatment. On the top panel we plot the evolution of the minimum oxygen level recorded in the tissue.}
	\label{fig:comp_int_RT}
\end{figure} 

In summary, our model suggests that the optimal schedules for UHFRT and FRT may differ markedly due to the different ways in which the composition of the tumours evolve during the two treatment protocols. For UHFRT, if doses are too close to each other, then treatment selects for CSCs which drive subsequent disease relapse. If, instead, the doses are too far apart, then the tumour has time to regrow, and to generate a severely hypoxic region, which reduces the overall treatment efficacy and drives treatment failure. By choosing an intermediate interval between doses (i.e., 1.5 weeks), the model predicts successful extinction of the tumour for physically realistic protocols (with total RT doses of 70 Gy). 
The absence of an hypoxic \emph{niche} in this case, in combination with the
conversion of potential CSCs into radio-sensitive DCs, enables
the radiation to eradicate the tumour.
By contrast, FRT becomes less effective as the time between treatment cycles increases. This is because, tumour re-oxygenation in FRT is less pronounced than in UHFRT, so that hypoxic \emph{niches} form more rapidly after the end of treatment. Consequently shorter intervals between treatment cycles are predicted to be more effective. Nevertheless, this approach leads to the early appearance of a larger fraction of CSCs and, thereafter, (under the conditions considered
here) to the survival of the tumour. 

\section{Conclusion}
\label{sec:conclusion}

We have proposed and analysed a new mathematical model, structured by phenotype and space, to describe the evolution of intra-tumour heterogeneity in the presence and absence of radiotherapy. We have focused on the role played by tissue oxygen levels in shaping the tumour's internal  phenotypic composition, with hypoxia (i.e., abnormally low oxygen levels) being a key driver of cell de-differentiation towards an aggressive, stem-like phenotype. 
The limited susceptibility of the CSCs to treatment with radiotherapy is found to be a key factor in driving the tumour's survival.

After introducing our model in \S\ref{sec:model}, we analysed its long time behaviour in the absence of treatment, combining analytical and numerical techniques to investigate its bifurcation structure as two model parameters vary: the magnitude of the differentiation velocity $V_+$ and the dimensionless oxygen (and, by extension,
nutrient) consumption rate $\gamma^*$ which measures the ratio of the timescales for oxygen diffusion and consumption i the tissue. We find that for biologically reasonable values of these parameters the system possesses two non-negative steady states which we term ``tumour-extinction'' and ``tumour-invasion''. Based on extensive numerical simulations, we conclude that relatively small tumours characterised by large values of $\gamma^*$ can successfully invade the tissue, while larger tumours with lower oxygen consumption rates $\gamma^*$ may eventually be eliminated. This suggests that the size of the basin of attraction of the ``tumour-extinction'' steady states decreases with $\gamma^*$. This result can be better understood by focusing on the internal composition of a tumour and establishing whether it contains sufficiently large hypoxic regions. Hypoxia niches are essential for the accumulation and persistence of cancer stem cells, which, due to their high clonogenic potential, are responsible for the growth and successful invasion of the tumour. 

In \S\ref{sec:dynamics_with_radio}, we have used the model to study tumour responses to radiotherapy (RT). We find that large values of $\gamma^*$ correlates with more aggressive tumours that respond poorly to treatment. This is mainly due to the difficulty in reoxygenating the tumour rather than the presence, prior to treatment, of radio-resistant CSCs. In particular, our model predicts that treatment will be successful only if the tumour burden at the end of treatment is low enough to prevent the formation of a durable hypoxic region after the end of treatment. However, as shown in \S\ref{sec:dyn_sim}, the maximum tumour burden that prevents relapse depends on the internal phenotypic composition of the tumour at the end of treatment and is, therefore, case (or patient) specific. When comparing different treatment protocols, we find that larger doses administered less often can be more effective than standard of care fractionation (2 Gy for 5 days per week). In \S\ref{sec:alternative_protocols}, we further investigate how different scheduling of the same dose fractionation protocol can impact the outcome of treatment. We find that, in certain cases, ``sooner'' is not always ``better'' and waiting longer between treatment cycles can be beneficial. This is because RT selects for radio-resistant phenotypes and therefore, if doses are too close to each other, resistant cells accumulate until eventually the tumour stops responding to treatment. At the same time, waiting too long can also be harmful, as the tumour has the time to grow sufficiently large and to develop extremely hypoxic regions where radio-resistant CSCs accumulates and all tumour cells are less radio-sensitive due to the lack of oxygen. 

A natural question is whether our conclusions generalise to more realistic, higher-dimensional spatial geometries, e.g., the framework of tumour spheroids~\cite{Bull2020}. When symmetry arguments are used to reduce 3D-problems to 1D geometries, we find that the qualitative behaviour of the solution is similar to that observed for the 1D-Cartesian problem investigated here (results not shown). However, fully 2D and 3D simulations may exhibit additional behaviours not captured by the simplified 1D-geometry. 

An interesting prediction of our model is that the rate of oxygen consumption by the cells plays a key role in determining treatment (here, radiotherapy) outcomes. This emphasises the possible benefit of combining RT and metabolic inhibitors to improve response to treatment, as recently proposed by several studies~\cite{Ashton2016,Fiorillo2016,Mudassar2020}. By decreasing the rate at which the cells consume oxygen, oxygen levels in the tumour can be increased to prevent the formation of regions of radio-biological hypoxia and increase the tumour's overall sensitivity to radiotherapy. Furthermore, our model suggests that metabolic inhibitors can also prevent the emergence of CSCs, by re-oxygenation of large hypoxic \emph{niches} in which stem-like cells initially might accumulate. A natural extension of the model would be to introduce the effect of metabolic inhibitors and investigate how they should be combined with RT to optimise treatment outcomes. Furthermore, when considering the use of high RT doses ($\geq 10$ Gy) in a clinical setting, RT side-effects should also be included into the model. 

While our model captures the higher radio-resistance of CSCs, recent findings suggest that radio-therapy can directly impact the differentiation status of cells \cite{Walcher2020}. In particular, RT-induced de-differentiation has been observed \emph{in vitro} and linked to the accumulation of senescent-type cells that can support the \emph{de novo} formation of CSCs by influencing the tumour micro-environment. Once more  information about the mechanism linking senescence and stemness become available, they could be easily incorporated into our model by for example allowing the phenotypic advection velocity itself (i.e., $v_s$) to depend on other environmental factors besides oxygen. Such additions would affect the evolution of the tumour heterogeneity during treatment, shifting the system into the basin of attraction of the ``tumour-invasion'' steady state solution. This raises the interesting question of what treatment strategies may be more effective when RT-induced radio-resistance is included and whether these may differ from the discussion herein.

Ultimately, one can envision enlarging the cells' phenotypic state space by introducing additional ``synthetic'' axes  that can capture more fully the cancer cells heterogeneity and evolution. However, the complexity of the model would make it difficult to gain insight into the role of the different ``phenotypic'' variables. Hence the need to start analysing simpler model, where a subset of these ``synthetic'' axes are included. From this point of view, given the amount of experimental evidence linking metabolism, stemness and treatment outcomes, a natural next step would be to include in the model two structure variables accounting for the stemness and metabolic status of cells. 
While previous theoretical studies have focused on the ways in which nutrient levels and/or treatment shape metabolic heterogeneity in tumours (e.g,  \cite{Ardaseva2020}, \cite{Hodgkinson2019},\cite{Villa2021}), the role of stemness is yet to be investigated. A model that captures all of these aspects could provide insight into the evolution of complex tumour ecosystem and help in the design of effective treatment strategies. 
While these goals are intriguing from a theoretical viewpoint, a different, but equally important, direction involves attempting to apply the model to experimental data. The development of RNA-seq techniques has resulted in a wealth of data being collected on the evolution of cells phenotypes (see, e.g.,~\cite{Patel2014}). The question then is how these rich datasets can be used to inform phenotypic-structured models. From this point of view, important practical aspects have to be addressed, such as parameter identifiability and estimation, that can be challenging when considering highly-parameterised and computationally-expensive models. Addressing these limitations will be crucial for translating our theoretical study into clinically-useful insight.

\section*{Funding}
\noindent~The authors wish to thank Prof Christos E. Zois for the helpful and illuminating discussions on the biology underpinning the model. G.L.C. is supported by from EPSRC and MRC Centre for Doctoral Training in Systems Approaches to Biomedical Science and Cancer Research UK. G.L.C. acknowledges also the support from Worcester College via the Martin Senior Scholarships. P.G.K. acknowledges support from the Leverhulme Trust via a Visiting Fellowship and thanks the Mathematical Institute of the University of Oxford for its hospitality during the early stages of this work.

\appendix
\section{Numerical methods for dynamical simulations}
\label{App:numerics}
We discretise the spatial variable $X$ using standard central finite difference on a stensil of $m_x+1$ points, where the Dirichlet condition in Eq.~(\ref{eq:c_stationary}) is strongly imposed, while ghost points are used to enforce the Neumann boundary conditions. For the phenotypic axis $s$, we use instead finite volume where we divide the axis into $m_s$ cells. Following~\cite{Gerisch2006}, we control the advection component using a Koren limiter. As discussed in~\cite{CELORA2021}, the use of a finite volume scheme allows us to properly account for the advective structural flux.
After having discretised both the $X$ and the $s$ axis, we obtain from Eq.~(\ref{eq:c_stationary}) a set of $m_x+1$ non-linear algebraic constraints ($\vec{g}(\vec{n}(t),\vec{c}(t))=0$) coupled to a system of $(m_x+1)\times m_s$ non-linear odes of the form $\dot{\vec{n}}(t)=\vec{f}(\vec{c}(t),\vec{n}(t),t)$ derived from Eq.~(\ref{eq:rescaled_sys}), where $\vec{n}(t)\in \mathcal{R}^{(m_x+1)m_s}$ and $\vec{c}(t)\in \mathcal{R}^{(m_x+1)}$. Note that, given $\vec{n}$, the implicit function $\vec{g}$ uniquely defines $\vec{c}$, i.e., we can write $\vec{c}=\vec{c}(\vec{n})$. We define $\tilde{\vec{f}}(\vec{n},t)=\vec{f}(\vec{c}(\vec{n}(t)),\vec{n}(t),t)$, so that the evolution of the vector function $\vec{n}(t)$ can be written as $\dot{\vec{n}}(t)=\tilde{\vec{f}}(\vec{n}(t),t)$. To advance in time, we use an implicit multistep \emph{bdf15} method for stiff equation from the \emph{scipy} library in python. Practically the input function $\tilde{\vec{f}}$ to the \emph{bdf15} method reads 
\begin{algorithm}
	\DontPrintSemicolon
	\SetKwFunction{FMain}{$\tilde{\vec{f}}$}
	\SetKwProg{Fn}{Function}{:}{}
	\Fn{\FMain{$\vec{n}$, $t$}}{
		Solve for $\vec{c}$ using fix point iterations\;
		Assemble the vector $\tilde{\vec{f}}$ using the computed oxygen distribution $\vec{c}$\;
		\KwRet\;
	}
\end{algorithm}

\noindent When including treatment, we want to properly capture the fast time scale of RT delivery. To do so, at each time $t_i$ at which a dose of RT is delivered, we integrate the solution ${\vec{n}}$ up to time $t_i$ using \emph{bdf15} method (as described above). In the interval $[t_i,t_i+\delta T]$, with $\delta T=0.5$, we instead use a simple explicit Euler scheme with small fixed time step $dt=5\times10^{-3}$; we then switch back to \emph{bdf15} method up to time $t_{i+1}$, when the next radiation dose is delivered. 

\section{Parameter values}
We summarise here the parameter values used in the simulation of Eqs.~(\ref{eq:rescaled_sys})-(\ref{eq:c_stationary}) presented in the main text. In Table~\ref{tab:parameters1}, we list the value of dimensional parameter values that could be estimated from the literature. The values of non-dimensional parameters are instead given in Table~\ref{tab:parameters2}.
\begin{table}[ht]
	\centering
	\caption{Summary of typical dimensional values for some of the parameters in Eq.~(\ref{eq:spatial_mod_no_scale}), togehter with supporting references.}
	\label{tab:parameters1} 
	\begin{tabular}{l@{\hspace{0.65cm}} l@{\hspace{0.65cm}} c}
		\toprule[1.5pt]\addlinespace[2pt]
		Parameter &  Reference value(s) & Refs\\\addlinespace[2pt]
		\hline\addlinespace[4pt]
		$D_{xn}$ &$10^{-10}-10^{-8}$ [mm$^2$/s] &\cite{Swanson2000,Villa2021}\\[3pt]
		$p^{max}_0$, $p^{max}_1$& 0.005, 0.02  [1/hr]& \cite{CELORA2021,Sweeney1998} \\[2pt]
		OER & 3 & \cite{Lewin_2020}\\
		$\alpha_{min}$, $\alpha_{max}$ &0.005, 0.1 [1/Gy]&\cite{Saga2019}\\
		$\beta_{min}$, $\beta_{max}$& 0.002, 0.01 [1/Gy$^2$]&\cite{Saga2019}\\
		$\gamma_0 C_\infty$ &$5–50 $ [mmHg/s] &\cite{Grimes2014}\\
		$D_{xc}$ & $2\times 10^{-3}$ [mm$^2$/s]&\cite{Grimes2014} \\[3pt]
		$C_\infty$ & $50$ [mmHg] & \cite{McKeown2014, West2019}\\[2pt]
		$C_H$ & $15$ [mmHg]&\cite{McKeown2014, West2019}\\[2pt]
		$C_{R}$ & $5$ [mmHg]&\cite{McKeown2014, West2019}\\[2pt]
		$C_N$ & $\approx 0$ [mmHg]&\cite{West2019} \\
		$L$ & $\approx 200 $ [$\mu$m]&\\
		\bottomrule[1.5pt]
	\end{tabular}
\end{table}
\begin{table}[ht]
	\centering
	\caption{Summary of non-dimensional parameters groups that appears in Eq.~(\ref{eq:rescaled_sys}). Parameter values are consistent with those reported in Table~\ref{tab:parameters1} and/or in \cite{CELORA2021}.}
	\label{tab:parameters2} 
	\begin{tabular}{l@{\hspace{0.65cm}} l@{\hspace{0.5cm}} }
		\toprule[1.5pt]\addlinespace[2pt]
		Parameter &  Value(s) \\\addlinespace[2pt]
		\hline\addlinespace[4pt]
		$\parD{sn}$ & $5\times 10^{-6}$ [1/hr] \\[3pt]
		$V_+$, $V_-$ & $8$, $2$ [$\times 10^{-4}$/hr]\\
		$\xi_+$, $\xi_-$& $0.1$, $0.1$ \\
		$\omega_+$, $\omega_-$& 1, 2\\[2pt]
		$\parD{xn}$&$ 10^{-4}$ [1/hr]\\[3pt]
		$K_{0}$, $K_{1}$& $0.05$, $0.3$\\[2pt]
		$g_0$, $g_1$ &0.01, 0.04 \\[2pt]
		$d_f$ &$0.001$ [1/hr] \\[2pt]
		$k_f$ & $10$   \\[2pt]
		$\xi_R$ & 0.2\\
		$\gamma^*$&$2-20$ [1/hr]\\
		$\parD{xc}$ & $180$ [1/hr]\\[3pt]
		$c_\infty$ & $1.0$\\[2pt]
		$c_H$ & 0.3\\[2pt]
		$c_R$ & 0.1\\[2pt]
		$c_N$ &0.0125\\
		\bottomrule[1.5pt]
	\end{tabular}
\end{table}
\clearpage
\section{Bifurcation Analysis}
\subsection{Linear stability analysis of the trivial steady state}
\label{App:LSA}
In this section, we determine the linear stability of the trivial steady state by seeking solutions of Eq.~(\ref{eq:rescaled_sys})-(\ref{eq:rescaled_sysOx1}) of the form:
\begin{equation}
n=\delta n e^{\lambda t}+O(\delta^2), \quad c=c_{\infty}+\delta ce^{\lambda t}+O(\delta^2),\label{eq:ansatz}
\end{equation}
where $0<\delta c,\delta n\ll 1$.
Substituting~(\ref{eq:ansatz}) into Eqs.~(\ref{eq:rescaled_sys})-(\ref{eq:rescaled_sysOx1}) and retaining terms which are linear in $\delta$ we obtain the following linearised problem:
\begin{subequations}
	\begin{align}
	\lambda \delta n=\parD{xn} \frac{\partial^2 \delta n}{\partial X^2}+\frac{\partial }{\partial s} \left(\parD{sn} \frac{\partial \delta n}{\partial s}-\delta n v_s(s,c_\infty)\right)+	\mathcal{F}(s,c_\infty,0)\delta n,\label{eq:pert_n}\\
	\lambda \delta c = \parD{xc}\frac{\partial^2 \delta c}{\partial X^2}-\gamma \parD{xc}\delta \phi,\label{eq:pert_ox}\\
	\parD{xs} \frac{\partial \delta n}{\partial s}-\delta n v_s = 0, \qquad s\in\left\{0,1\right\},\, X\in [0,1],\, t>0,\\
	\left.\frac{\partial \delta c}{\partial X}\right|_{X=0}=0, \quad \delta c(1,t)=0, \quad t>0,\\[2pt]
	\left.\frac{\partial \delta n}{\partial x}\right|_{X=0}=\left.\frac{\partial \delta n}{\partial x}\right|_{X=1}=0, \quad s\in(0,1),\, t>0.
	\end{align}\label{sys:perttot}%
\end{subequations}
where $\delta \phi(X)=\int_0^1 \delta n(X,s) ds$. We note that Eq.~(\ref{eq:pert_ox}) for $\delta c$ decouples from Eq.~(\ref{eq:pert_n}) for $\delta n$.
We can therefore solve Eq.~(\ref{eq:pert_n}) independently of $\delta c$, and seek a separable solution of the form:
\begin{equation} 
\delta n = \Lambda(x)W(s)
\end{equation}
where $\Lambda$ and $W$ needs to satisfy:
\begin{subequations}
	\begin{align}
	\begin{aligned}
	\lambda =\frac{\parD{xn}}{\Lambda} \frac{d^2\Lambda}{d X^2}+ \frac{1}{W}\frac{d }{d s} \left(\parD{sn} \frac{d W}{d s}-  v_sW\right)
	+ (p(s,c_\infty)-f(s)),\end{aligned}\label{eq:full_op}\\
	\parD{sn} \frac{d W}{d s}= 0, \qquad s\in\left\{0,1\right\}, \label{app:cond1}\\[2pt]
	\left.\frac{d \Lambda}{d X}\right|_{X=0}=\left.\frac{d \Lambda}{d X}\right|_{X=1}=0,
	\end{align}\label{sys:sep_var_pert}
\end{subequations}
where in writing Eq.~(\ref{app:cond1}) we have used the fact that the velocity $v_s$ vanishes at the boundaries.

Using standard arguments, we find that the eigenfunctions $\Lambda_m$ associate to the eigenvalue $\omega_m$ are given by
\begin{subequations}
	\begin{align}
	\Lambda_m= \cos\left(\omega_m X\right), \quad \omega_m= \pi m, \quad m=0,1,2,\ldots.
	\end{align}
	so that the eigenfunctions $W$ associated to the phenotypic axis, $s$, must satisfy the following ODE:
	\begin{align}
	\left(\omega_m^2 \parD{xn} +\lambda\right) W = \frac{d }{d s} \left(\parD{sn}  \frac{d W}{d s}-  v_sW\right)
	+\left[p(s,c_\infty)-f(s)\right]W\equiv \mathcal{M}(W),\\
	\parD{sn} \frac{d W}{d s}= 0, \qquad s\in\left\{0,1\right\}.
	\end{align}
\end{subequations} 
Here, the operator $\mathcal{M}$ is equivalent to the differential operator obtained by linearising the well-mixed formulation of Eqs.~(\ref{eq:rescaled_sys}), which is obtained by setting all spatial derivatives to zero and considering an homogeneous oxygen concentration. The spectrum of the operator $\mathcal{M}$ has been investigated in more detail in \cite{CELORA2021}. While we have no explicit solution for the eigenvalues $\kappa_k$ and eigenfunctions $W_k$ of $\mathcal{M}$, we can show that $\kappa_k$ are real and bounded from above.
If $\parD{xn}$ is constant and $W_k$ denotes the eigenfunctions of $\M$ associated to the eigenvalue $\kappa_k$, then $\delta n_{m,k} =\Lambda_m(X)W_k(s)$ is an eigenfunction of the operator, Eq.~(\ref{eq:full_op}), with real eigenvalue $\lambda_{m,k} =\kappa_k -\omega_m^2 \parD{xn}$. Since $\omega^2 \parD{xn}\geq 0 $, we find that $\lambda_{m,k}\leq \kappa_k$ and $\lambda_{m,k}<\lambda_{m',k}$ if $m'<m$. Consequently, $\lambda_{m,k}< 0$ for all $k$ and $m$ if and only if $\bar{\kappa}=\max_k\kappa_k<0$. 

The question now is whether $\lambda=\bar{\kappa}$ is an eigenvalue for the full system Eqs.~(\ref{sys:perttot}). If this is the case, then the stability of the trivial steady-state for the model Eqs.~(\ref{eq:rescaled_sys})-(\ref{eq:rescaled_sysOx1})
boils down to studying the linear stability of the steady state for the well-mixed model for homogeneous oxygen concentration $c_\infty$.

Let us consider the case $\omega_m\equiv 0$ for which wlog $\Lambda_0 \equiv 1 $. In this case, the eigenfunction associated to the eigenvalue $\bar{\kappa}$ is $\delta n=W_{\bar{\kappa}}(s)$. The eigenfunction is defined up to a multiplicative constant which we set so that $\int_0^1 W_k(s)=\delta \phi\equiv \Delta>0$. In order for $\bar{\kappa}$ to be an eigenvalue of Eqs.~(\ref{sys:perttot}), Eq.~(\ref{eq:pert_ox}) must have a solution for $\lambda=\bar{\kappa}$ and $\delta \phi\equiv\Delta$. Substituting for $\lambda$ and $\delta \phi$, we obtain:
\begin{subequations}
	\begin{align}
	\bar{\kappa} \delta c +\gamma^* \parD{xc} \Delta = \parD{xc}\frac{\partial^2 \delta c}{\partial X^2},\\ 
	\delta c (1)= 0, \quad \partial_x \delta c(0)=0,
	\end{align}
	which has the solution:
	\begin{align}
	\delta c=\bar{C}(X)=\begin{cases} \frac{\gamma^* \parD{xc}\Delta}{\bar{\kappa}}\left[\frac{\cosh(\bar{\tau} X)}{\cosh(\bar{\tau} )}-1\right], &\quad \bar{\kappa}>0,\\
	\frac{\gamma^*\Delta}{2} (X^2-2), &\quad \bar{\kappa}=0,\\
	\frac{\gamma^* \parD{xc}\Delta}{\bar{\kappa}}\left[\frac{\cos(\bar{\tau} X)}{\cos(\bar{\tau} )}-1\right], &\quad \bar{\kappa}<0,
	\end{cases}
	\end{align}
	where $\bar{\tau}=\sqrt{|\bar{\kappa}/\parD{xn}|}$.
\end{subequations}
Given that a solution exists, we have that $\bar{\kappa}$ belongs to the spectrum of the full operator, Eqs.~(\ref{sys:perttot}), with eigenfunction $(\bar{W}(s),\bar{C}(X))$. Further, since the spectrum of Eqs.~(\ref{sys:perttot}) is a subset of $\lambda_{m,k}\leq \bar{\kappa}$, $\bar{\kappa}$ will be the eigenvalue with largest real part and therefore dictates the stability of the trivial solution. 
From~\cite{CELORA2021}, we know that $\bar{\kappa}$ decreases from positive to negative values as the magnitude of the advection velocity $V_+$ increases. The critical value $V_+^{(cr)}$ at which $\bar{\kappa}=0$, corresponds to a transcritical bifurcation. This bifurcation must therefore occur also in the spatial model, and, based on the above result, the location of the transcritical bifurcation $V_+=V_+^{cr}$ is independent of $\gamma^*$, $\parD{xn}$ and $\parD{xc}$. We conclude therefore that the critical value $V_+^{(cr)}$ at which the trans-critical bifurcation occurs will not been affected by the inclusion of spatial effects, in line with our
numerical observations in the text.

\subsection{Numerical continuation and stability estimates}
In order to compute the equilibrium states $(n_{\infty}(x,s),c_{\infty}(x))$ of the full system~(\ref{eq:rescaled_sys})-(\ref{eq:rescaled_sysOx1}) in the absence of treatment, we simply set to zero all time derivatives to obtain:
\begin{subequations}
	\begin{align}
	E_n(n_{\infty},c_{\infty})=0, \quad E_c(n_{\infty},c_{\infty})=0,\quad \vec{E}_{bc}(n_{\infty},c_{\infty})=\vec{e}_6,
	\end{align}
	where $\vec{e}_6=[0,0,0,0,0,1]^T$ and
	\begin{align}
	E_n(n,c)&=\parD{xn} \frac{\partial^2 n}{\partial X^2}+ \frac{\partial }{\partial s} \left(\parD{sn} \frac{\partial n}{\partial s}-v_s(s,c)n\right)+ \mathcal{F}(s,c,\phi) n,\\
	E_c(n,c)&=\parD{xc}\frac{\partial^2 c}{\partial X^2}-\gamma^*\parD{xc} \phi H_{\epsilon_{\gamma}}(c-c_N),\\[6pt]
	\vec{E}_{bc}(n,c)&=\left[ \left.\frac{\partial n}{\partial s}\right|_{s=0},  \left.\frac{\partial n}{\partial s}\right|_{s=1},\left.\frac{\partial n}{\partial X}\right|_{X=0},\left.\frac{\partial n}{\partial X}\right|_{X=1}, \left.\frac{\partial c}{\partial X}\right|_{X=0},\left.c\right|_{X=1}\right].
	\end{align}\label{eq:def_equi}%
\end{subequations}
Due to the non linearities in the system, we cannot solve for Eqs.~(\ref{eq:def_equi}) explicitly. We instead rely on numerical approaches in particular continuation techniques to investigate the system equilibria and their stability. Let us consider a perturbation of the equilibrium points:
\begin{equation}
n=n_{\infty}+e^{\lambda t}\delta n, \quad  c=c_{\infty}+e^{\lambda t}\delta c,\label{eq:pert}
\end{equation}
where the perturbations are small, i.e. $|\delta c|,|\delta n|\ll1$, and the growth rate $\lambda \in \mathbb{C}$ is a complex number.
\begin{subequations}
	Substituting Eqs.~(\ref{eq:pert}) into the system~(\ref{eq:rescaled_sys})-(\ref{eq:rescaled_sysOx1}), we obtain:
	\begin{align}
	\lambda \begin{bmatrix}
	\delta n\\
	\delta c
	\end{bmatrix} &=J(n_{\infty},c_{\infty})\begin{bmatrix}
	\delta n\\
	\delta c
	\end{bmatrix},\\[4pt]
	J(n_{\infty},c_{\infty})&=\begin{bmatrix}
	\frac{\delta E_n}{\delta n}(n_{\infty},c_{\infty}) &\frac{\delta E_n}{\delta c}(n_{\infty},c_{\infty})\\[5pt]
	\frac{\delta E_c}{\delta n}(n_{\infty},c_{\infty}) & 	\frac{\delta E_c}{\delta c}(n_{\infty},c_{\infty})\end{bmatrix}
	\end{align}
	with the constraint:
	\begin{align}
	\vec{E}_{bc}(\delta n,\delta c) =\vec{0}.
	\end{align}\label{eq:stab}%
	The stability of the steady state can therefore be related to the spectrum of the Jacobian $J$ in particular $(n_{\infty},c_{\infty})$ will be stable if $\Re(\lambda)<0$ for all $\lambda$ such that Eq.~(\ref{eq:stab}) admits a nontrivial solution.
\end{subequations}

We solve the nonlinear problem~(\ref{eq:def_equi}) numerically via parameter continuation techniques using \texttt{Julia}'s package \texttt{BifurcationKit}~\cite{veltz2020}, which allows us to draw the bifurcation diagram for the system as a function of a given parameter. To do so we discretise Eqs.~(\ref{eq:def_equi}) as described in~\ref{App:numerics} and we compute the discretised approximation of $E_n,E_c$ and $J$. By computing the eigenvalues of the matrix $J$ we can approximate the eigenvalues of the differential operator~(\ref{eq:stab}) and study the stability of the computed steady states. To characterise the solution we define the total number of cells in the domain:
\begin{equation}
M_{\infty}= \int_0^1 \phi_\infty(X) dX.
\end{equation}

\subsection{Fold continuation}
\label{app:foldcont}
\begin{figure}[htb]
	\centering
	\includegraphics[width=0.9\textwidth]{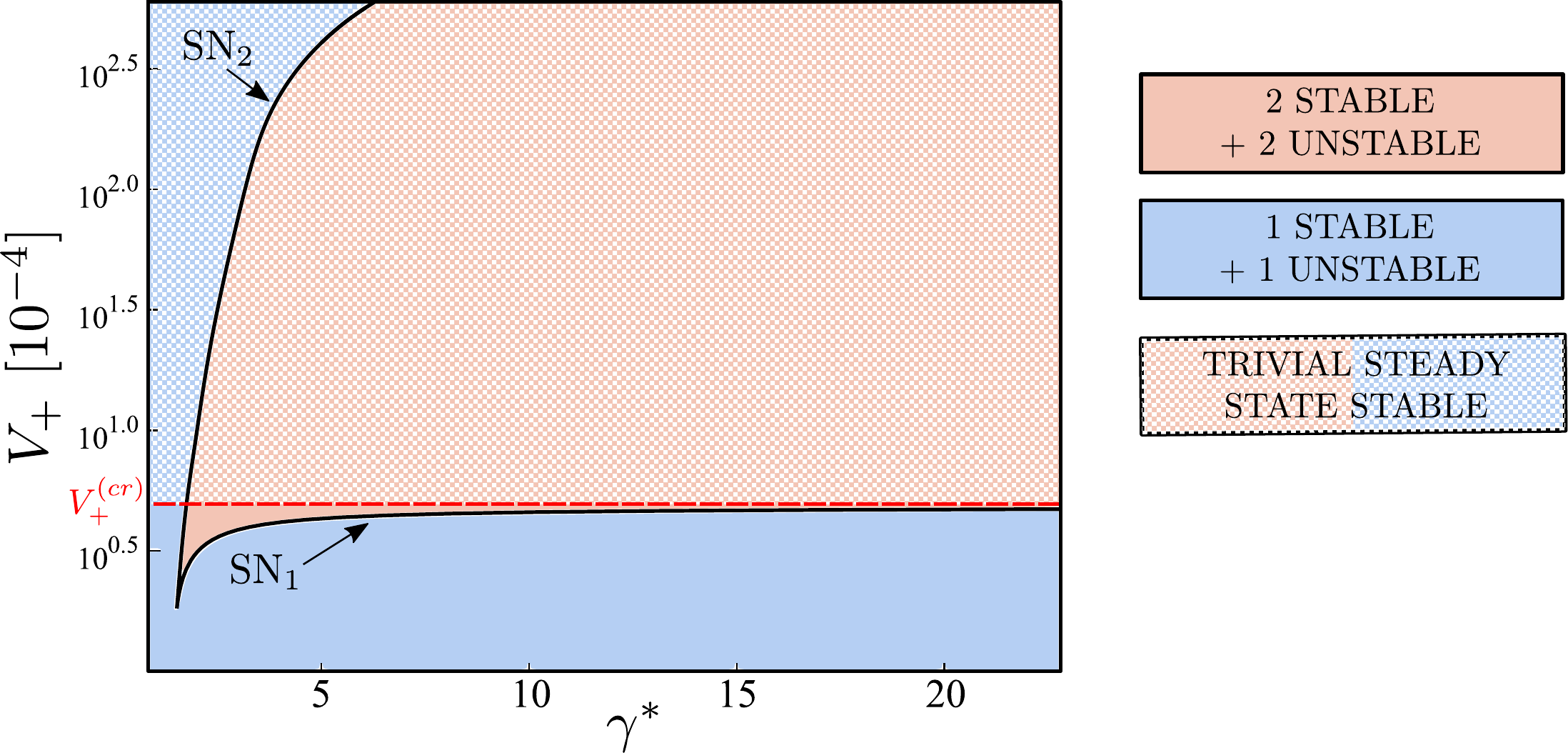}
	\caption{Continuation of the fold points $\sn{1}$ and $\sn{2}$ in the ($\gamma^*$,$V_+$) parameter space. Here $\sn{1}$ and $\sn{2}$ are as identified in the bifurcation diagrams in Figs.~\ref{fig:bif_diag1} and \ref{fig:bif_diag2}. The points  $\sn{1}$ and $\sn{2}$ divide the parameter spaces in regions where Eqs.~(\ref{eq:def_equi}) admits a different number of solutions (as indicated by the different colours). The two saddle nodes originates from a cusp point.}
	\label{fig:foldcont}
\end{figure} 

Using \texttt{Julia}'s package \texttt{BifurcationKit}~\cite{veltz2020}, we are also able to compute the location of saddle-node bifurcation points and follow their location as we vary the values of model parameters. This, however, is highly computationally expensive as it augments the system~(\ref{eq:def_equi}), with an additional $(m_x+1)m_s$ constraints to define the fold points. We were therefore constrained in the number of points $m_x$ and $m_s$ that we could use to discretise the system. This implies that the results are not highly accurate but can still be informative in understanding the bifurcation structure trends of the system under study. 

Looking at Fig.~\ref{fig:foldcont}, we see that for sufficiently small $\gamma^*$, there are no fold points (i.e., the bifurcation diagram is analogous to Fig.~\ref{fig:bifwellmix}). The two fold points $SN_1$ and $SN_2$ originates from a singular cusp (i.e., a codimension two) 
point, and then depart from each other. While $SN_2$ steeply increase reaching large values of $V_+$, the trajectory of $SN_1$ asymptotes to a finite value of the advection velocity, $V_+^{(cr)}$, as $\gamma^*\rightarrow \infty$. Here $V_+^{(cr)}$ corresponds to the value of $V_+$ for which the system has a transcritical bifurcation (see Fig.~\ref{fig:bif_diag1}). As discussed in~\ref{App:LSA}, the location of $V_+^{(cr)}$ is independent of the parameter $\gamma^*$ and divides the $(\gamma^*,V_+)$ plane into two regions: the one where the trivial steady state is stable (see white dotted region) and the one where it is unstable.

\bibliographystyle{elsarticle-harv} 
\bibliography{biblio}

\end{document}